\documentclass[twocolumn]{aastex631}

\usepackage{array}
\usepackage{amsmath}
\usepackage{makecell}
\usepackage{multirow}

\usepackage{hyperref}
\newcommand{\eparam}{\boldsymbol{\theta}}

\newcommand{\eparamdi}{\boldsymbol{\theta}_i}

\newcommand{\D}{\boldsymbol{d}}

\newcommand{\DDi}{\boldsymbol{d}_i}

\newcommand{\hyp}{\boldsymbol{\Lambda} }
\newcommand{\NOBS}{N_{\rm obs} }

\newcommand{\trig}{{\rm tr} }

\newcommand{\model}[1]{\emph{#1}}

\newcommand{\IMRELEASENO}{LLNL-JRNL-2004065-DRAFT}

\newcommand{\llnl}{Space Science Institute, Lawrence Livermore National Laboratory, 7000 East Ave., Livermore, CA 94550, USA}
\newcommand{\UCB}{University of California, Berkeley, Astronomy Department, Berkeley, CA 94720, USA}
\newcommand{\UCR}{Department of Physics and Astronomy, University of California, Riverside, 900 University Avenue, Riverside, CA 92521, USA}
\newcommand{\UM}{Department of Physics, University of Michigan, 450 Church St, Ann Arbor, MI 48109, USA}
\newcommand{\LCTP}{Leinweber Center for Theoretical Physics, 450 Church St, Ann Arbor, MI 48109, USA}

\begin{document}

\title{Hints of an Anomalous Lens Population towards the Galactic Bulge}

\author[0000-0002-5910-3114]{Scott E. Perkins\,*}
\email{*perkins35@llnl.gov}
\affiliation{\llnl}
\author[0000-0002-1052-6749]{Peter McGill\,}
\affiliation{\llnl}
\author[0000-0003-0248-6123]{William A. Dawson\,}
\affiliation{\llnl}
\author[0000-0002-4457-890X]{Ming-Feng Ho\,}
\affiliation{\UCR}
\affiliation{\UM}
\affiliation{\LCTP}
\author[0000-0002-0287-3783]{Natasha S. Abrams}
\affiliation{\UCB}
\author[0000-0001-5803-5490]{Simeon Bird\,}
\affiliation{\UCR}
\author[0000-0001-9611-0009]{Jessica R. Lu}
\affiliation{\UCB}

\begin{abstract}
The dark and dynamic parts of the Galaxy, including the bulk shape and movement of the Galactic Bulge and characteristics of dark compact object populations, such as a hypothetical population of primordial black holes (PBHs), are difficult to study directly by their very nature, but are critical to our understanding of the universe. 
Fortunately, all of these mysteries can be uniquely studied via gravitational microlensing, a method of astronomical detection that traces mass and dynamics as opposed to light.
Using the OGLE-IV microlensing survey bulge fields, we apply a Bayesian hierarchical model to jointly infer properties of the Galaxy, the characteristics of compact objects, and and test PBHs with an extended mass distribution as a test PBHs as a viable explanation of dark matter, extending work focused on the Small and Large Magellanic Clouds, both with much lower event-rates.
We infer a preference within the data for a lower patternspeed in the galactic model and a wider mass spectrum for compact objects.
When adding a PBH component to the favored astrophysical model from our initial investigations, we find a Bayes factor of $\ln\mathcal{B} = 20.23$ preferring the PBH model. 
Upon further investigation of these results, we find the critical feature in the PBH model to be the velocity distribution, which is fundamentally different than the velocity distribution of astrophysical objects and uniquely able to explain a large number of low parallax, low timescale microlensing events.
Noting that this effect is not unique to PBHs, we consider the implications of these results as applied to a hypothetical population of PBHs and discuss alternative explanations, including a variety of other possible astrophysical and survey or analysis systematics.
\end{abstract}


\section{Introduction}\label{sec:intro}

There are many mysteries about the Universe which are fundamentally difficult (or even impossible) to study with traditional observation methods.
For example, the bulk properties of the Galactic bulge, including its spatial configuration, mass function and dynamics, has been a difficult problem to study, as our Milky Way Galaxy can only be studied from our perspective within the galaxy itself~\citep[for example, see][]{Robin2012}.
Additionally, the characteristics of isolated dark objects, like stellar origin black holes (SOBHs) and most neutron stars (NSs), are difficult to study directly, as they do not produce electromagnetic or gravitational radiation in any appreciable amount.
Primordial black holes (PBHs) lie at the intersection of these two problems in many ways: they are individually dark and impossible to detect via radiation when in isolation, and are only discernible from normal SOBHs when considering their properties in the context of population models (i.e., PBHs should have different distributions in space, mass and velocity than SOBHs)\footnote{While true for certain mass ranges, this is less strict for other mass ranges. There are no known mechanisms to create a compact object below roughly a solar mass, for example.}.
These PBHs also have cosmological significance as a possible explanation of dark matter (DM), causing them to experience a resurgence in the last few years~\citep{Bird2023}.

In contrast to most observational techniques in astronomy, gravitational microlensing probes objects' \emph{mass and dynamics}, not their light. 
This positions microlensing to be a very effective probe in the context of each of these mysteries~\cite{Rose2022,Lu2019,Wyrzykowski2020,Gould2000,Lam2022,Sahu2022,Sweeney2022,Sweeney2024,Perkins2024,Wegg2017}.
In fact, a prime motivator for microlensing science, historically, has been to investigate DM, and in particular, to study Massive Compact Halo Objects (MACHOs)~\cite{Alcock2001, Tisserand2006, Blaineau2022,Wyrzykowski2009,Wyrzykowski2010,Wyrzykowski2011a,Wyrzykowski2011b}, of which PBHs are a subset, as a possible explanation. 
Now, with decades of consistent observations available, one can investigate these questions of galactic dynamics, compact object formation physics and even the possibility of PBHs in unprecedented detail.

As PBHs are particularly well suited to be studied with microlensing, we consider them as a specific addition to the astrophysical model when investigating the galactic model and compact object physics.
They are predicted to have formed via density fluctuations in the early universe \citep{Zeldovich1967,Hawking1971}. Constraints on the PBH population could provide insights into the nature of dark matter \citep{Chapline1975}, the seeds of supermassive black holes \citep[e.g.,][]{Bernal2018}, and the physics of the early universe \citep[e.g,][]{Carr1975}. However, this dark and isolated population of black holes has eluded definitive detection over the last 50 years of investigation \citep[e.g.,][]{Green2021}.

\begin{figure}
    \centering
    \includegraphics{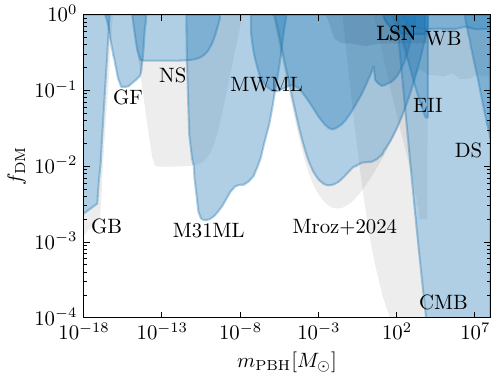}
    \caption{Current PBH constraints (blue and gray regions indicate more and less conservative assumptions, respectively), assuming a monochromatic mass spectrum at $m_{\rm PBH}$ taken from \cite{Bird2023}. M3ML and MWML are derived from microlensing towards M31 \citep{Niikura2019M31} and looking through the Milky Way, respectively \citep{Alcock2001, Tisserand2006, Wyrzykowski2011b, Blaineau2022}. Mroz+2024 are the recent microlensing constraints towards the LMC from \citep{Mroz2024}. Other constraints are supernovae lensing \citep[LSN;][]{Zumalacarregui2018},  Eridanus II dwarf galaxy \citep[EII;][]{Brandt2016,Li2017}, wide binary stars \citep[WB;][]{Quinn2009,Yoo2004}, dwarf galaxy dynamical heating \citep[DH;][]{Lu2021,Takhistov2022, Takhistov2022b}, X-ray binaries \citep[XB;][]{Inoue2017}, CMB distortions \citep[CMB;][]{Ali-Haimoud2017, Ricotti2008}, disk stability \citep[DS;][]{Xu1994}, gamma-ray background \citep[GB;][]{Carr2016}, gamma ray-femtolensing \citep[GF;][]{Carr2016dm}, and neutron star capture \citep[NS;][]{Capela2013}.}
    \label{fig:current constraints}
\end{figure}

The main challenge in determining whether a population of isolated PBHs could explain some fraction of dark matter is that there is no means of direct detection. Instead, the various effects that PBHs are expected to have on their surroundings are used to place limits on the fraction of dark matter mass that can be comprised of PBHs ($f_{\rm DM}$). Fig. \ref{fig:current constraints} shows the current constraints on PBHs in $f_{\rm DM}$ -- PBH mass space ($m_{\rm PBH}$) assuming a monochromatic mass spectrum. For $m_{\rm PBH}\gtrsim10^{3}M_{\odot}$, distortions in the Cosmic Microwave Background rule out even a small $f_{\rm DM}$ \citep[][]{Ricotti2008, Ali-Haimoud2017}. At the lower-mass end, $m_{\rm PBH}\lesssim10^{-17}M_{\odot}$, the gamma ray-background constrains $f_{\rm DM}\lesssim10^{-2}$ \citep{Carr2016}. In the intermediate mass-ranges, $10^{-10}M_{\odot}<m_{\rm PBH}<10^{3}M_{\odot}$, the most competitive constraints on $f_{\rm DM}$ come from gravitational microlensing surveys \citep{Alcock2001, Tisserand2006, Wyrzykowski2011b, Niikura2019M31,Blaineau2022, Mroz2024} and constrain $f_{\rm DM}\lesssim10^{-3}-10^{-1}$ with similar constraints coming from quasar microlensing~\citep{Esteban-Gutierrez:2023qcz,Mediavilla:2017bok}.
Relaxing these assumptions is expected to result in less restrictive constraints.

Despite the superior sensitivity of microlensing to compact objects in the intermediate mass ranges, robustly interpreting this probe is challenging \citep[e.g.,][]{Wyrzykowski2009, Niikura2019ogle}. These challenges arise because a single photometric microlensing event does not contain direct information on the characteristics of the lensing object. Instead statistics over many microlensing events and event rates are used with a model of the Galaxy to infer $f_{\rm DM}$ \citep[e.g.,][]{Alcock2001}. Typically in these analyses, a set of restrictive assumptions is used: a monochromatic PBH mass spectrum is assumed, and a fixed phenomenological Galactic model is used to derive $f_{\rm DM}$.

To mitigate against not being able to tell if any given event is caused by a PBH, the strategy employed to infer $f_{\rm DM}$ by the first microlensing surveys \citep[][]{Alcock1993,Udalski1993,Aubourg1993,Muraki1999} was to observe through the outer regions of the Galactic halo, where for any appreciable $f_{\rm DM}$ the dominant source of microlensing events should be compact objects \citep[][]{Paczynski1986}.
This strategy of focusing on regions of low total event rate, but also of low lens confusion, continues to be popular \citep[e.g.,][]{Niikura2019M31,Blaineau2022}. 
However, because $f_{\rm DM}$ is apparently low, looking through the Galactic Halo ends up still suffering from the same lens confusion problem - stellar self-lensing in the Magellanic Clouds \citep{Sahu1994, Wyrzykowski2009, Wyrzykowski2011} and a potential free-floating planet (FFP) population \citep{Niikura2019ogle} have both hindered definitive and robust $f_{\rm DM}$ constraints.  

Nearly all previous $f_{\rm DM}$ constraints using microlensing observations have assumed a monochromatic PBH mass spectrum.
This assumption is common across all PBH probes \citep[e.g.,][]{Carr2016dm,Zumalacarregui2018,Li2017,Takhistov2022b} and permits constraints to be compared in the $f_{\rm DM}-m_{\rm PBH}$ space shown in Fig. \ref{fig:current constraints}.
Despite the monochromatic mass spectrum's straightforward interpretation and computational simplicity, it is restrictive and not necessarily representative of a physical PBH mass spectrum which could be more extended \citep[e.g.,][]{Clesse2015,Carr2016,Inomata2017}. Relaxing the monochromatic mass spectrum assumption can change derived $f_{\rm DM}$ constraints \citep{Carr2017, Calcino2018} and provide insight into the apparent tensions between dynamical and microlensing constraints on a PBH population \citep{Green2016}.
While studies using gravitational waves (GWs) to study PBHs have examined the possibility of extended PBH mass functions \citep[e.g.,][]{Franciolini2021, Wong2020}, no previous microlensing study has attempted to jointly and agnostically infer $f_{\rm DM}$ and the shape of an extended PBH mass spectrum.

Microlensing constraints on $f_{\rm DM}$ are always understood as the residual against a Galactic model of the astrophysical population of lenses -  Stars, White Dwarfs (WDs), Neutron Stars (NS), and Stellar Origin Black Holes (SOBHs).
The majority of previous microlensing analyses have assumed a fixed Galactic model (and a model of the Magellanic clouds) with the implicit assumption that the abundance and properties of the astrophysical lens population are known perfectly.
This assumption is more reasonable for the well-studied population of luminous lenses (Stars and White Dwarfs) than it is for the population of isolated black holes, which are expected to be abundant \citep[e.g.,][]{Samland1998, Olejak2020} but only one has been detected \citep{Sahu2022, Lam2022, Lam2023}.
Moreover, modeling the Galactic model jointly with $f_{\rm DM}$ is key to disentangling a PBH dark matter component from the Stellar population and understanding systematic sources of confusion in microlensing observables \citep{Calcino2018, Perkins2024}.

The current state of the art microlensing analyses in the context of PBH constraints are \cite{Mroz2024} and \cite{Calcino2018}. \cite{Mroz2024} used two different models of the Galaxy to infer $f_{\rm DM}$ leading to a maximum difference in the upper limit of $f_{\rm}$ at the $\approx1\%$ level for $m_{\rm PBH}\approx1M_{\odot}$.
While the methods of \cite{Mroz2024} bookend some sources of Galactic model uncertainty, the Galactic models are still fixed and could therefore underestimate effects of systematic confusion between the Galactic model and a PBH dark matter population. \cite{Calcino2018} showed that different models of the dark matter halo and their measured uncertainties could significantly weaken and even remove microlensing constraints looking towards towards the Magellenic clouds for dark matter with masses $1M_{\odot}-10 M_{\odot}$. 

In this work we infer both $f_{\rm DM}$ and the shape of an extended PBH mass spectrum using microlensing data from Optical Gravitational Lensing Experiment towards the Galactic Bulge \citep[OGLE;][]{Udalski2003, Udalski2015}.
Although the Galactic Bulge line of sight benefits from $\approx 100$-times higher microlensing event rate \citep[e.g.,][]{Udalski1994,Mroz2019, Mroz2024optical} when compared with looking through the Galactic Halo towards the Magellanic Clouds, stellar lenses dominate the event numbers for any $f_{\rm DM}$, which means that this dataset has not been used to constrain $f_{\rm DM}$ thus far.

To meet the challenge of inferring the characteristics of a putative dark matter PBH population using microlensing events towards the Galactic Bulge, addressing deficiencies of assuming a monochromatic mass, and accounting for some sources of Galactic model uncertainty, we use an adapted version of the hierarchical Bayesian framework presented in \cite{Perkins2024}.
This framework treats the class of any given lens probabilistically, allowing $f_{\rm DM}$ to be inferred while accounting for the fact that we do not know the class of any individual lens with any certainty.
This framework also allows an arbitrarily shaped PBH mass spectrum to be inferred jointly with $f_{\rm DM}$ and simulation-based Galactic model.
Similar methods were applied in the context of GW data~\cite{Franciolini2021}, but with very different sets of benefits and systematics.
GWs probe extragalactic populations, preferentially investigating BHs in dense environments (in order to form binaries which merge in a Hubble time).
Furthermore, both microlensing and GW need to disambiguate events as being astrophysical or cosmological, but GWs must also contend with the uncertainty of the merger rate for PBHs in order to infer fundamental physics of the underlying theory.
This is not a consideration for the microlensing efforts presented here.

Such flexibility in modelling and the relaxation of the assumed Galactic model opens our analysis up to catching all types of systematic effects, whether they were caused from a missing cosmological population (such as PBHs), missing astrophysical populations (e.g., Brown Dwarf stars, BDs, free floating planets, FFPs, or binary systems), mischaracterization of the efficiencies of the OGLE survey (such as suggested in ~\citealt{Mroz2019}) or systematics due to individual lightcurve modeling and inference~\citep[see for example][showing the need for more sophisticated modeling]{Golovich2022}.
Because of this, the results of this analysis must be viewed through this lens of possible mis-modeling: any residual signal in the data attributed to a PBH population could just as likely be caused by one of these known systematics. 
Therefore, we will also discuss these results and the implication of the residual signal in our inference in the context of these known astrophysical contexts.

This paper is organized as follows. We cover the basics of microlensing and data used in this study in Sections \ref{sec:microlensing} and \ref{sec:data}, respectively. We then detail our inference framework and population model in Section \ref{sec:framework}. Finally, we present our results in Section \ref{sec:results} and discuss them in the context of current and potential future PBH constraints in Section \ref{sec:conclusions}.

\section{Microlensing}\label{sec:microlensing}

\noindent During a photometric microlensing event, the lens with mass $M_{L}$ at distance $D_{L}$ deflects the light of a more distance background source at distance $D_{S}$. During this close chance alignment between the lens and source two unresolved images of the source, which change in brightness, are formed, causing an apparent transient amplification of the source flux by \citep{Paczynski1986}:
\begin{equation}
A(t) = \frac{u(t)^2 + 2}{u(t)\sqrt{u(t)^2 + 4}}\,.
\end{equation}
Here $u$ is the magnitude of lens-source angular separation normalized by the angular Einstein radius $\theta_E = \sqrt{4 G M c^{-2} \left(D^{-1}_L - D^{-1}_S\right)}\,$. 
Two parameters that contain information on the lens-source distance and velocity and the lens mass can be constrained from the photometric microlensing signal. The first is the Einstein crossing-time,
\begin{equation}\label{eq:tEDef}
t_{E}=\frac{\theta_{E}}{\mu_{\rm rel}},
\end{equation}
where $\mu_{\rm rel}$ is the relative lens-source proper motion. The second is the microlensing parallax,
\begin{equation}
    \pi_{E} = \frac{\pi_{rel}}{\theta_{E}}; \quad \pi_{rel} = 1
\text{AU}\left(D^{-1}_{L} - D^{-1}_{S}\right),
    \label{eq:pi_E}
\end{equation}
which is constrained via the annual parallax effect via a an accelerating on-earth observer or simultaneous observation from space~\citep{Refsdal1966,Zhu2017,McGill2023} and causes asymmetrical deviations in the microlensing lightcurve \citep{Alcock1995}. Putting this all together, the observed flux during a microlensing event is,
\begin{equation}
F(t; \boldsymbol{\theta}) =  F_{\rm Base} + b_{\rm sff}F_{\rm Base}\left[A(t; u_{0}, t_{0}, t_{E}, \pi_{E}, \phi) - 1\right].
\label{eq:flux}
\end{equation}
Here, $F_{\rm Base}$ is the total baseline flux including flux from the source, lens, and unrelated neighbors, and $b_{\rm sff}$ is the fraction of source light to $F_{\rm Base}$, $\phi$ is the angle between the ecliptic north and the direction of the lens-source relative proper motion in the heliocentric frame (east of north). 
The impact parameter in units of $\theta_E$ is denoted as $u_0$, and $t_0$ represents the time of closest angular approach.
This leaves $\boldsymbol{\theta}=\{F_{\rm Base},b_{\rm sff}, t_0, t_E, u_0, \pi_{E}, \phi \}$ as the full set of parameters that describes a microlensing event.
In what follows, we use the annual microlensing parallax parameterization in \cite{Golovich2022}.

\begin{figure}
    \centering
    \includegraphics[width=\linewidth]{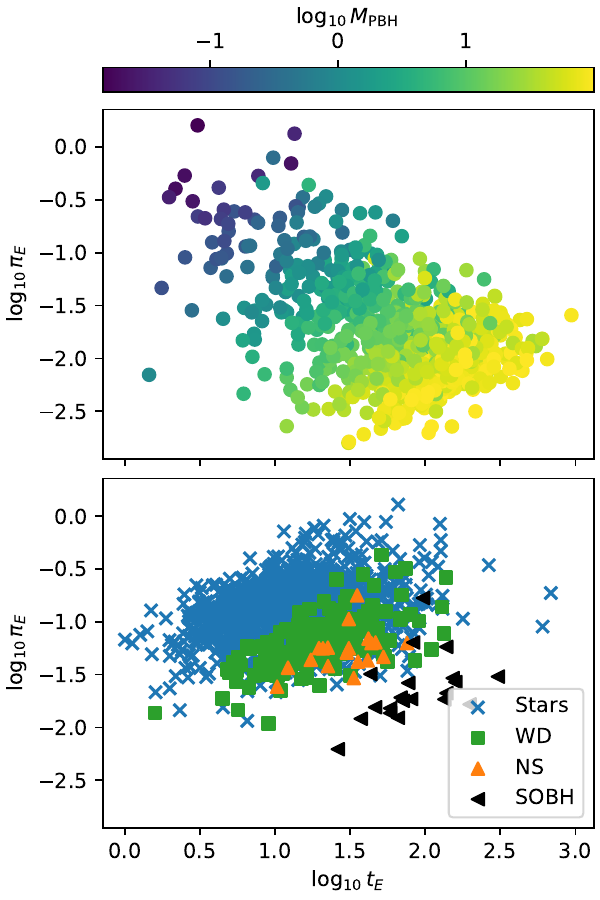}
    \caption{Simulation of the joint distribution of $t_{E}$ and $\pi_{E}$ for several OGLE bulge fields using \texttt{PopSyCLE} \citep{Lam2020} with the PBH modifications from \cite{Pruett2024}. The top panel shows the PBH lenses, which were injected with a true mass distribution log-uniform from $10^{-2}M_{\odot}$ to $10^{2}M_{\odot}$, colored by their masses. The lower panel shows the astrophysical lens populations. The different subpopulations of lenses occupy different, but overlapping, areas of the $t_{E}$-$\pi_{E}$ space. 
    The simulations were processed to ensure all microlensing events satisfy $|u_0| < 1$ and $I < 21$, as discussed in  Sec.~\ref{sec:forwardModeling}.
    }
    \label{fig:pruett_sim}
\end{figure}

In this scenario, for a single event, even if both $t_{E}$ and $\pi_{E}$ are well constrained from the photometric signal, direct lens information such as mass and distance cannot be extracted.
However, if a large sample of events are detected over the course of a survey, characteristics of the lens population can be constrained \citep[e.g.,][]{Mroz2017, Sumi2023}.
Lens population characteristics can be constrained because different types of lenses (e.g. Stars, BHs, White Dwarfs) have different kinematics, mass spectra, and spatial distributions in the Galaxy and therefore cause microlensing events with different $\pi_{E}$ and $t_{E}$ \citep{Lam2020}.
Fig. \ref{fig:pruett_sim} shows the distributions of $\pi_{E}$ and $t_{E}$ for different lens types from a Population Synthesis Code for Compact Object Microlensing Events \citep[\texttt{PopSyCLE};][]{Lam2020} simulation with an injected PBH dark matter population \citep{Pruett2024} spanning a range of masses PBHs.
The different lens types occupy different but overlapping regions of the $t_{E}-\pi_{E}$ space, allowing the different population of lenses to be characterized using these observables \citep{Perkins2024}.

\section{Data}\label{sec:data}

We use a set of 5970 I-band photometric microlensing events analyzed in \cite{Golovich2022} from the fourth \citep[OGLE-IV][]{Udalski2015} phase of the OGLE survey. Specifically, we used the lightcurves of the 5970 events analyzed and made publicly available in \citep{Mroz2019}. \footnote{We omit the 2617 events detected in the OGLE high-cadence regions of the bulge presented in \citep{Mroz2017} due to their lightcurve data not being public. We also do not include the 630 OGLE events found in the Galactic plane and presented in \cite{Mroz2020}.} The sample of microlensing events considered in the work were detected between 2010–2017 in the ongoing OGLE-IV phase and have $\approx 1<t_{E}<100$ days.

Alongside the lightcurve data we use the $t_{E}$ event detection efficiency curves from \cite{Mroz2019} (hereafter $p_{\rm det}(t_E)$). The $t_{E}$ efficiency curves were computed via catalog-level simulations \citep[e.g.,][]{Afonso2003, Wyrzykowski2009}, and defined between Einstein crossing-times between 1 and 316 days. 
Below this range, we fix the efficiency at its value at one day, and above this range, we set the detection efficiency to zero.
For OGLE-IV, \cite{Mroz2019} computed efficiency curves for each OGLE field containing microlensing events to capture the effects of the heterogeneous observing strategy over the survey area. All efficiency curves were obtained from the OGLE data download site\footnote{\url{http://ogle.astrouw.edu.pl/}}.

Finally, we use the posterior inference on the microlensing events parameters obtained by \cite{Golovich2022} by fitting a microlensing model (including parallax effects) jointly with a Gaussian process model that accounted for systematic instrumental effects and unlensed stellar variability.

\section{Lens Population Model}\label{sec:framework}

\subsection{Choosing Observables}

This first step in the analysis is to decide which characteristics of a single event should be used to make inferences on the lens population. In microlensing, typically the first observable considered is $t_E$ \citep[e.g.,][]{Sumi2023,Mroz2017,Allsman2001,Tisserand2006,Blaineau2022}. 
$t_{E}$ is the natural choice as it usually well-constrained \citep[within $\approx16\%$;][]{Mroz2019} and traces lens mass ($t_{E}\propto\sqrt{M}$; Eq. \ref{eq:tEDef}). This makes $t_{E}$ a good discriminator between different lens subpopulations (e.g. Stars, WDs, NSs and BHs) because they span $\sim4$ orders of magnitude in mass. Indeed, assuming that relative lens-source distances and velocities average out over a survey, $t_{E}$ is frequently considered a surrogate for the lens-mass \citep[e.g.,][]{Griest1991,Allsman2001,Tisserand2006,Blaineau2022}.
However, this is only useful as general guidance, as the Einstein crossing-time is also dependent on the spatial configuration and relative velocity of the microlensing source and lens, via $\pi_{\text{rel}}$  and $\mu_{\text{rel}}$.

In addition to $t_{E}$, $\pi_{E}$ also contains information on the less mass ($\pi_E \propto 1/\sqrt{M}$; Eq.~\ref{eq:pi_E}), and can be measured by an accelerating on-earth observer or near-simultaneous observations separated in space.
While $\pi_{E}$ has been useful for the analysis of single events \citep[e.g.,][]{Lam2022,Sahu2022}, it is typically only well-constrained for a small subset of events \citep[e.g.,][]{Wyrzykowski2020,Kaczmarek2022}, and therefore is not usually considered in microlensing populations studies. 
However, Fig. \ref{fig:pruett_sim} shows that there is structure and discriminating power in the joint $t_{E}-\pi_{E}$ distribution for different lens classes \citep{Lam2020}, so even a weak constraint (i.e., an upper limit) on $\pi_{E}$ may provide some population information.
An additional parameter that could contain some information on the type of lensing object is $b_{\rm sff}$ - nearby events caused by dark lenses should be systematically less blended than events caused by luminous lenses \citep[e.g.,][]{Wyrzykowski2016}.
However, we found that \texttt{PopSyCLE} simulations show that for the majority of events this effect is washed out by random unresolved third-light resulting in no significant separation in $b_{\rm sff}$ for lens classes.
We therefore use the joint $t_{E}-\pi_{E}$ observable space in this analysis.

\subsection{Hierarchical Framework}\label{sec:stats}

We use an adapted version of the Bayesian hierarchical framework developed in \cite{Perkins2024}. This framework takes into account survey selection bias, event rate Poisson statistics, and lens classification uncertainty. The framework is based on established hierarchical models \citep[e.g.,][]{Loredo2004,Mandel2018,Taylor2018,Vitale2020} and is adapted for use with photometric microlensing surveys. For a detailed explanation and derivation of this method see Sec. 3 and Appendix A of \cite{Perkins2024}, \cite{Kaczmarek2025} and \cite{Sallaberry2024}. Here, we briefly outline the method's key components. 

The posterior distribution on the population hyperparameters, $\hyp$, which contain information on the lens populations (e.g., $f_{\rm DM}$, the PBH mass spectrum, and various parameters of the Galactic model) we want to infer is
\label{eq:popPost}
\begin{equation}\label{eq:hyperpost}
p(\hyp | \{\DDi\},\NOBS) \propto p(\hyp)e^{-\alpha N(\hyp)}N(\hyp)^{\NOBS}  \prod_{i=0}^{\NOBS}\mathcal{L}_i^{{\rm obs}}\,.
\end{equation}
Here, $\DDi$ is the lightcurve of a single microlensing event and $\{\DDi\}$ is the set of all microlensing lightcurves. $\NOBS$ is the total number of observed microlensing events over the survey, $N(\hyp)$ is the number of expected microlensing events from the population model, and $p(\hyp)$ is the prior on the population hyperparameters, which we detail in Section~\ref{sec:priors}. $\alpha$ describes a population model's efficiency at producing detectable events under a microlensing survey's selection function and is given by the average probability to observe an event (conditioned on said population model):%
\begin{equation}\label{eq:alpha}
\alpha = \int p_{\rm det} (t_E) p( t_E | \hyp)d t_E \,.
\end{equation}
Here, $p_{\rm det}(t_E)$ is the OGLE survey efficiency curves as a function of $t_{E}$ published by the OGLE collaboration \citep{Wyrzykowski2015,Mroz2019}. Finally, $\mathcal{L}_i^{{\rm obs}}$ is the likelihood of a single event under the population model and is given by
\begin{equation}\label{eq:popEventLikelihood}
\mathcal{L}_i^{{\rm obs}}  = \int p(\DDi|\eparamdi)p(\eparamdi|\hyp)d\eparamdi \,.
\end{equation}
$\eparamdi$ are the microlensing parameters for the ith event defined in Eq. (\ref{eq:flux}).  
$p(\eparamdi| \hyp)$, sometimes called the population distribution or the population-informed prior, comes from our population forward modeling, detailed in Section~\ref{sec:forwardModeling}. 
The event likelihood $p(\DDi|\eparamdi)$ describes the probability of seeing data $\DDi$ given it was produced by an event with microlensing parameters $\eparamdi$ and we use the noise model defined in Eq. (27) of \cite{Golovich2022}. Eq. (\ref{eq:popEventLikelihood}) is an integral over all $\boldsymbol{\theta}_i$. 
Therefore, by only considering the $t_{E}$ and $\pi_{E}$ distributions of the population models, we are implicitly assuming there are no discriminating features between the distributions of all the other single event parameters between the different lens subpopulations. 
We define the hyperparameters, $\hyp$, used in our population model in Section \ref{sec:forwardModeling}.

Under the assumption that the number and distributions of microlensing events caused by each different class of lens (classes = \{Stars, NS, WD, SOBH, PBH\}) are independent, we can write the population likelihood of a single event as a mixture model:
\begin{equation}\label{eq:mixmodel}
p(\eparam_{i} | \hyp)  = \sum_{\text{class}\in \text{classes}} \psi_{\rm class} p(\eparam_{i} | \text{class}).
\end{equation}
Here $\psi_{\rm class}$ is the intrinsic relative abundance of each lens class, and we define the vector of these abundances as $\boldsymbol{\psi}=[\psi_{\rm Star}, \psi_{\rm WD}, \psi_{\rm NS}, \psi_{\rm SOBH}, \psi_{\rm PBH}$], where each of these fractions are free parameters in our model.
For the astrophysical subpopulations, we numerically estimate $p(\eparam_{i} | \text{class})$ by fitting a kernel density estimate \citep[using \texttt{SciPy;}][]{Virtanen2020} in $\pi_{E}-t_{E}$ space to the samples of each lens class, produced by the procedure outlined in Section~\ref{sec:forwardModeling}.
For the case of the PBH distribution, which spans almost 4 orders of magnitude in mass, we implement a more flexible model described in Section \ref{sec:mass_distribution_modeling}.
By using this model, we achieve some level of flexibility in the astrophysical components, absorbing any systematics in the Galactic model by inferring the relative abundances which determine the overall impact of each subpopulation.
However, by still training the kernel density estimators on the samples directly (rather than using a fully non-parametric model), we retain a physical interpretation, critical for connecting our inference back to fundamental science.

\subsection{Modeling residual signals}\label{sec:mass_distribution_modeling}

On top of the astrophysical models outlined above, we will also consider two flexible subpopulations to assess any residual information in the data, not captured by the astrophysical modeling (i.e., \texttt{PopSyCLE}).
We will consider two types of models, one meant to model a subpopulation of PBH lenses while the other reflects a subpopulation of astrophysical lenses. 
Both models will be described with a similar framework, described below. 

\begin{figure}
    \centering
    \includegraphics[width=\linewidth]{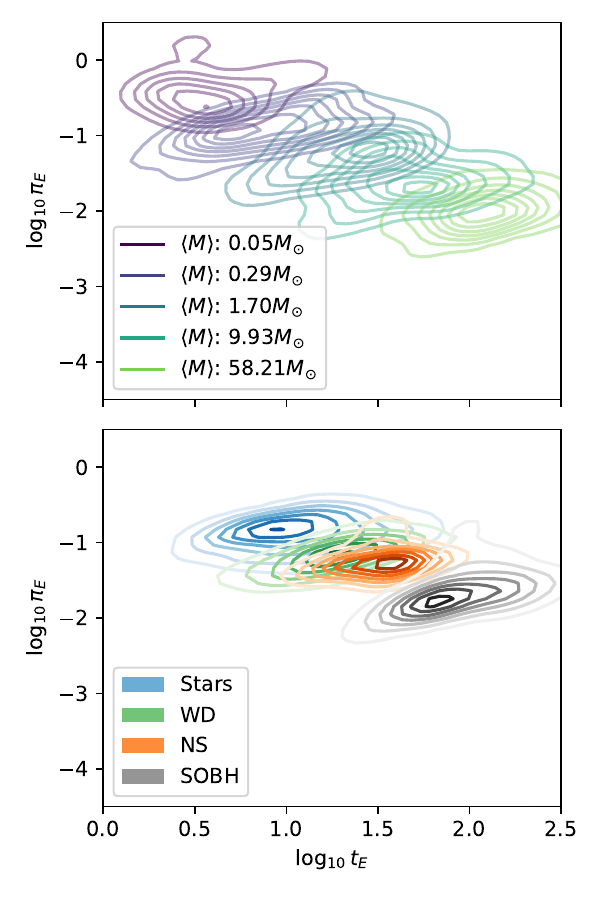}
    \caption{
    A visualization of the predicted distributions of observables, separated by different types of lenses, is shown above.
    The top panel shows the distributions for each of the different five (logarithmically spaced) mass bins used to model the PBH population.
    The total PBH representation in this space is a mixture model constructed from linear combinations of these distributions.
    The bottom panel shows the astrophysical distributions, for reference.
    }
    \label{fig:5bin_pbh_model_observable_space}
\end{figure}

To model the shape of this residual lens subpopulation mass spectrum, we use a flexible step function, \citep[see e.g.,][for the application of this method to the population $t_{E}$ distribution]{Golovich2022, Kaczmarek2024} defined as
\begin{equation}\label{eq:pbh_mass_distribution}
    p_{\boldsymbol{\beta}}(\log_{10}M)=\sum_{b=1}^{N_{B}}\frac{\beta_{b}}{\Delta \log_{10} M}\,s\left(\frac{\log_{10} M}{\Delta \log_{10} M };\frac{b-1}{N_{B}} ,\frac{b}{N_{B}}\right),
\end{equation}
where $\boldsymbol{\beta}\equiv \{\beta_1 , \ldots \beta_{N_B}\} $ parameterizes the mass spectrum and $\sum_{b=1}^{N_{B}} \beta_b = 1$.
The quantity $\Delta \log_{10} M$ represents the total range in mass allowed by the model, and $N_B$ represents the number of bins.
The functions $s(x;L,H)$ are defined as 
\begin{equation}
    s(x; L, H) = 
    \begin{cases}
    0 & \text{for } x < L, \\
    (H - L)^{-1} & \text{for } L\leq x\leq H,\\
    0 & \text{for } x > H
    \end{cases}
\end{equation}
where we divide the allowed range of masses for PBHs into 5 bins (logarithmically spaced) between $10^{-2} M_{\odot}$ and $10^{2} M_{\odot}$.
For the astrophysical lenses, we similarly divide the space into 5 bins (logarithmically spaced) between the minimum and maximum lens mass in the underlying astrophysical simulation.
The number of bins were chosen through exploratory analysis such that the overlap in the observable space ($t_E$-$\pi_E$) was minimal to avoid an overly complex model, but sufficiently high to achieve accurate reconstruction of a general mass distribution to within the expected resolution scale by the measurement accuracy of a ground-based microlensing survey.
The different distributions for the PBH model are shown in Fig.~\ref{fig:5bin_pbh_model_observable_space}.

This model can then be connected with the distributions of observable parameters through marginalization. Starting by considering the quantity of ultimate interest ($p(t_E , \pi_E | \text{class} )$ for $\text{class} \in \{\text{PBH}, \text{Astrophysical residual}\}$), we can relate this to the distribution we just defined as follows,
\vspace{.5mm}
\begin{widetext}
\begin{align}\label{eq:pbh_observable_distribution}\nonumber
   p(t_E, \pi_E | \text{class}  ) &= \int p(t_E, \pi_E, \log_{10} M| \text{class}) d\log_{10} M \,,\\\nonumber
    &= \frac{\Delta \log_{10} M}{N_{B} } \sum_b p(t_E, \pi_E| \log_{10} M, \text{class} ) p_{\boldsymbol{\beta}}(\log_{10}M) \,, \\
    &= \sum_b \beta_b p(t_E, \pi_E| \log_{10} M, \text{class} )  \,.
\end{align}
\end{widetext}
We have inserted the actual form of our mass distribution into this equation (Eq.~\eqref{eq:pbh_mass_distribution}) and evaluated the integral using the fact that our mass distribution is uniform in each bin (simplifying the integral into a sum).
The quantity $p(t_E, \pi_E| \log_{10} M, \text{class} )$ represents the probability of observable parameters $t_E$ and $\pi_E $, conditioned on the lens being either a PBH or astrophysical lens and the mass being $\log_{10} M$. 
This distribution is calculated by taking the simulation samples using the procedure outlined in Section~\ref{sec:forwardModeling}. 
For the PBH residual model, we bin the events involving a PBH lens into the different mass bins, then train a kernel density estimator \citep[using \texttt{SciPy;}][]{Virtanen2020} on the samples in each bin separately. 
The same is done for the astrophysical model, where the samples used to train the kernel density estimator in each bin are the (unlabeled in class) astrophysical lenses falling in said bin. 
While this binning procedure is an approximation, it is consistent with our original mass distribution in Eq.~\eqref{eq:pbh_mass_distribution}, which is agnostic to scales below the bin width. 
The key difference between these two models is the base assumptions used to map a bin in mass to the distribution in observable parameters $t_E$ and $\pi_E$. 
While the frameworks have similar form and are both parameterized by a flexible mass distribution, this mapping is the critical difference in their ability to describe the data.
The astrophysical residual model is sensitive to systematics in the mass distribution, while the PBH residual model allows for flexiblity in the mass distribution \emph{and} leads to predictions in $t_E$ and $\pi_E$ with different shapes and extents, due to the PBH samples having different velocity and spatial distributions.

\subsection{Population Forward modelling}\label{sec:forwardModeling}

To connect parameters of the lensing populations to the characteristics of observed events (i.e., to compute $p(\eparamdi|\hyp)$, $p( t_E | \hyp)$, and $N(\hyp)$ in Section \ref{sec:framework}) we used \texttt{PopSyCLE} with the PBH additions of \cite{Pruett2024}. \texttt{PopSyCLE} allows for the simulation of a microlensing survey given a model of the Galaxy. The underlying stellar population in \texttt{PopSyCLE} is  based on the the Besançon model \citep{Robin2004} implemented by \texttt{Galaxia} \citep{Sharma2011}.

SOBHs, NSs, and WDs are generated and injected into the simulation by evolving clusters matching each subpopulation of stars generated by \texttt{Galaxia} (thin and thick disk, bulge, stellar halo), which are then binned into approximately single-age and single-metallicity bins, which are then passed on to the Stellar Population Interface for Stellar Evolution and Atmospheres code \citep[\texttt{SPISEA};][]{Hosek2020} as single-metallicity clusters. 
\texttt{SPISEA} assumes an initial mass function, stellar multiplicity, extinction law, metallicity-dependent stellar evolution, and an a set of initial final mass relations \citep[IFMR; see e.g.,][]{Kalirai2008, Lam2020, Rose2022}. SOBHs and NSs are assigned initial natal kick velocities.
Furthermore, we also incorporate binaries in the simulation for one specific model (see Ref.~\citet{Abrams:2025} for details).

PBHs are distributed according to a NFW Halo density profile with parameter estimates from \cite{McMillan2017}. PBHs velocities are assigned according to an Eddington inversion model \citep{Lacroix2018} which is then used to define a Maxwellian distribution from which velocities are drawn. PBH velocities are also restricted to be less than the Milky Way escape velocity \citep[$v_{\text{esc}}<550$kms$^{-1}$;][]{Piffl2014}. Overall, this allows fast sampling of PBH velocities, but neglects correlations between PBH mass, location, and velocity \citep{Pruett2024}. All the main adopted \texttt{PopSyCLE} parameters and relationships are detailed in Table \ref{tab:flexibleHyperparameters} and Table~\ref{tab:galaxia_models}. Notably, \texttt{PopSyCLE} does not contain a brown dwarf (BD) or a free-floating planet (FFP) population. In each \texttt{PopSyCLE} simulation we used $20$ fields of radius $0.3$deg$^{2}$ detailed in Table \ref{tab:sim_fields}. Each simulation across all fields takes between $7000$-$17,000$ core-hours depending on the input simulation parameters.

\begin{table}[]
    \centering
    \begin{tabular}{c|c|c}
    \hline
    \hline
    l [$^{\circ}$] & b [$^{\circ}$] & area [deg$^{2}$]\\
    \hline
    \hline
    1.1399 & -3.7432 & 0.3 \\
    3.6341 & 2.1945 & 0.3 \\
    -4.2223 & -6.8055 & 0.3 \\
    5.8168 & 2.1491 & 0.3 \\
    6.6383 & -4.8152 & 0.3 \\
    3.5176 & 3.5577 & 0.3 \\
    4.4046 & -3.2761 & 0.3 \\
    -4.21 & 4.9609 & 0.3 \\
    4.6747 & 2.8534 & 0.3 \\
    3.3316 & -3.8823 & 0.3 \\
    0.7819 & 1.6875 & 0.3 \\
    -4.2794 & -5.4419 & 0.3 \\
    5.6025 & 4.8747 & 0.3 \\
    -3.2058 & -4.8329 & 0.3 \\
    5.4762 &-2.6684 & 0.3 \\
    -3.2832 &-3.4735 & 0.3 \\
    -1.0641 & -3.6101 & 0.3 \\
    -0.9534 &-6.3377 & 0.3 \\ 
    \hline
    \hline
    \end{tabular}
    \caption{Simulation fields used the \texttt{PopSyCLE} simulation. Shown above are the galactic coordinates (latitude and longitude) as well as the area of each simulation field.
    }
    \label{tab:sim_fields}
\end{table}

While the \emph{intrinsic} population of events is the quantity needed to evaluate Eq.~\eqref{eq:hyperpost} (i.e., the distribution of events observed or not), we must still conform to the decisions made by the OGLE collaboration in their assessment of the detection efficiency curves~\citep{Mroz2019}\footnote{See Appendix~C of Ref.~\citet{Perkins2024} for a detailed explanation of why this consistency in definition is important.}.
That is, the assumed distributions on $u_0$ and $I$ are implicitly encoded in the calculation of $p_{\rm det} (t_E)$, and we must therefore calculate our population models consistently. Therefore, we cut the simulation output to match that of OGLE: $|u_0| < 1$ and the source magnitude $I < 21$.

Similarly, the binary model also requires additional filtering before the simulation data is consistent with OGLE's definition of the intrinsic population. 
In the case of binaries, all microlensing events with multiple peaks are removed from the output, in addition to filtering on $|u_0|$ and $I$, mentioned above. 
Given OGLE did not model these binary populations when assessing detection efficiency, one can only marginalize over these extra dimensions in parameter space if the binary artifacts in the signal are perturbations on top of a PSPL signal.
Assuming the microlensing signal is PSPL-like or that a PSPL model would fit the signal well, ensures that using OGLE's detection efficiency is valid. 
Mathematically, this assumption amounts to $p_{\text{det}}(t_{E,\text{binary}}) \approx p_{\text{det}}(t_{E,\text{PSPL}})$, where $t_{E,\text{binary}}$ ($t_{E,\text{PSPL}}$) is the estimate of $t_E$ obtained using the system mass of the binary.

\begin{table*}
    \centering
    \begin{tabular}{l|p{75mm}||p{60mm}}
        \hline \hline
         Hyperparameter & Description & Prior/\{Options\}/Fixed Value   \\ 
         \hline \hline 
         $N$ & Total number of microlensing events (observed or not) over the survey duration & Uniform(lower=1,upper=$10^{5}$) \\
         
         $\boldsymbol{\beta}_{\rm PBH}$ & PBH mass spectrum bin heights & Dirichlet(concentration=$\boldsymbol{1}$) \\
         $\boldsymbol{\psi}$ & Intrinsic relative abundance of lens classes $\boldsymbol{\psi}=[\psi_{\rm Star}, \psi_{\rm WD}, \psi_{\rm NS}, \psi_{\rm SOBH}, \psi_{\rm PBH}$] &  Dirichlet(concentration=$\boldsymbol{1}$)\\
         $\boldsymbol{\psi}_{\neg \text{PBH}}$ & Intrinsic relative abundance of lens classes (neglecting PBHs) $\boldsymbol{\psi}_{\neg \text{PBH}}=[\psi_{\rm Star}, \psi_{\rm WD}, \psi_{\rm NS}, \psi_{\rm SOBH}$] &  Dirichlet(concentration=$\boldsymbol{1}$)\\
    IFMR & Initial-Final Mass Relation & \{\citet{Raithel2018}, \citet{Sukhbold2014}, \cite{Spera2015}\} \\
    $v_{\text{esc}}$ & Milky Way escape velocity & $550$kms$^{-1}$ \citep{Piffl2014} \\
    $r_{gc}$ & Sun-Galactic center distance  & $8.3$kpc \\
    $v_{\mathrm{kick, BH}}$ & Peak of initial SOBH progenitor kick distribution & $100$kms$^{-1}$ \\
    $v_{\mathrm{kick, NS}}$ & Peak of initial NS progenitor kick distribution  & $\{150$kms$^{-1},200$kms$^{-1},250$kms$^{-1},350$kms$^{-1}\}$ \\
    
    $\rho_{0}$ & Characteristic central density of the dark matter halo  & $0.0093M_{\odot}$pc$^{-3}$ \citep{McMillan2017} \\ 
    $\gamma$ & Slope parameter of the dark matter density profile &  1 (NFW; \citealt{Navarro1996})\\
    $r_{s}$ & Milky Way dark matter halo scale radius &  $18.6$kpc \\
    - & Extinction Law & \citet{Damineli2016} \\
    - & Singles or Binaries & \{Single systems only, binaries included\} \\
    - & Galactic Bulge Model (\texttt{Galaxia} Model)& \{ v${}_1$,v${}_2$,v${}_3$\}\\
         \hline \hline 
    \end{tabular}
    \caption{Description of parameters used for the \texttt{PopSyCLE} forward model. Free parameters which are inferred jointly in the model have a prior distribution specified. Parameters with  reference values are held fixed. Parameters with several discrete options that are explored in this work are enclosed in \{\}. The last row represents different options for the galactic model,  as proposed by ~\cite{Lam2020}. The relevant parameters that change between the models are outlined in Table~\ref{tab:galaxia_models}.
    This entire hyperparameter space is not explored simultaneously, but instead explored systematically.
    The astrophysical models we explore are outlined in Table~\ref{tab:astrophysical_models}.
    }
    \label{tab:flexibleHyperparameters}
\end{table*}

\begin{table*}[]
    \centering
    \begin{tabular}{c|p{40mm} | p{35mm} | p{35mm} | p{30mm} }
    \hline \hline
    Moniker &  Bar dimensions (radius, major axis, minor axis, height) &Bar angle (Sun–Galactic center, 2nd, 3rd) &Bulge velocity dispersion (radial, azimuthal, z) &  Bar patternspeed \\ \hline \hline
    v${}_1$ &(2.54, 0.70, 0.424, 0.424) kpc &(62.0, 3.5, 91.3)${}^{\circ}$ &(100, 100, 100) km s$^{-1}$ & 40.00 km s$^{-1}$ kpc${}^{-1}$\\  
    v${}_2$ &(2.54, 1.59, 0.424, 0.424) kpc &(78.9, 3.5, 91.3)${}^{\circ}$ & (100, 100, 100) km s$^{-1}$& 40.00 km s$^{-1}$ kpc${}^{-1}$\\  
    v${}_3$ &(2.54, 1.59, 0.424, 0.424) kpc & (78.9, 3.5, 91.3)${}^{\circ}$&(110, 110, 100) km s$^{-1}$ &78.62 km s$^{-1}$ kpc${}^{-1}$ \\  
    \hline \hline
    \end{tabular}
    \caption{Shown above is a summary of the different \texttt{Galaxia} models used in this article, meant to marginalize over uncertainties in modeling the Galactic Bulge itself. 
    The choices for the parameters follow those suggested and considered in \cite{Lam2020}.}
    \label{tab:galaxia_models}
\end{table*}

\begin{table}
    \centering
    \begin{tabular}{c|c | p{2cm} |p{2cm} }
    \hline\hline
    
    Moniker &IFMR & Galactic Bulge \newline Model & NS (SOBH) \newline kick velocity  (km s${}^{-1}$)  \\ \hline\hline
      \model{SukhboldN20}  &  SukhboldN20 & v3 &  350 (100)  \\
      \model{Spera15}  &  Spera15 & v3 &  350 (100)  \\
      \model{Raithel18}  &  Raithel18 & v3 &  350 (100)  \\
      \model{Galaxia v${}_1$}  &  SukhboldN20 & v1 &  350 (100)  \\
      \model{Galaxia v${}_2$}  &  SukhboldN20 & v2 &  350 (100)  \\
      \model{NS Kick 250}  &  SukhboldN20 & v3 &  250 (100)  \\
      \model{NS Kick 200}  &  SukhboldN20 & v3 &  200 (100)  \\
      \model{NS Kick 150}  &  SukhboldN20 & v3 &  150 (100)  \\
      \model{Binaries}  &  SukhboldN20 & v3 (binaries) &  350 (100)  \\\hline\hline
    \end{tabular}
    \caption{
    Models used to describe the astrophysical lensing populations.
    We consider our ``base'' model to be \model{SukhboldN20}, with each subsequent model testing a different assumption underlying said model.
    \model{Spera15} and \model{Raithel18} investigate IFMR choices, while \model{Galaxia v${}_1$} and \model{Galaxia v${}_2$} question galactic model choices.
    The three \model{NS Kick} models shift the mean of the Maxwellian distribution used to assign NS kick velocities by the stated amount.
    Finally, \model{Binaries} incorporates binary effects into source and lens systems.
    See Table~\ref{tab:flexibleHyperparameters} and Table~\ref{tab:galaxia_models} for details on these choices.
    }
    \label{tab:astrophysical_models}
\end{table}

\subsection{Parameters, Priors and Inference}\label{sec:priors}

We consider two classes of models: one only using the astrophysical lens populations and one including both PBHs and the astrophysical lens populations. 
When considering the astrophysical model, the free parameters of our model are the total number of microlensing events (observed or not) over the OGLE surveys and the relative abundances of each lens class neglecting the PBH subpopulation i.e. $\hyp=\{N, \boldsymbol{\psi}_{\neg \text{PBH}} \}$.
When considering the full model, including the PBH subpopulation, the free parameters are the bin heights of the PBH mass spectrum, the total number of microlensing events (observed or not) and the relative abundances of each lens class i.e. $\hyp=\{N,\boldsymbol{\beta}, \boldsymbol{\psi} \}$. 
We place a uniform prior on $N$ bounded between $1$ and $10^{5}$ that is designed to be diffuse and uninformative. 
We place a Dirichlet prior with concentration equal to the unit vector on $\boldsymbol{\beta}$, $\boldsymbol{\psi}$ and $\boldsymbol{\psi}_{\neg \text{PBH}}$ to enforce the required normalizations. 

Many sources of uncertainty in the Galactic model are treated in a discrete nature, either because the different models are naturally discrete or for computational tractability.
These models include the choice of IFMRs, the Galactic bulge orientation, and the average of the Neutron Star natal kick velocity distribution. 
The full list of \texttt{PopSyCLE} parameters, including the different choices for the above model components, is shown in Table~\ref{tab:flexibleHyperparameters}.

Examining Table~\ref{tab:flexibleHyperparameters} shows that there are $144$ possible discrete model combinations (2 for PBH/no PBH, 4 for NS kicks, 3 for Galactic model options, and 2 for singles/binaries, and 3 for IFMRs) to explore. Running the Galactic simulations, let alone the inference, for all of these models would cost up to $\approx2.4$ million core-hours and is therefore computationally intractable. 

Looking for any residual in microlensing data (e.g., a PBH population at a low $f_{\rm dm}$) first requires understanding the dominant signal of Stars, WDs, NSs, and SOBHs.
Therefore, to navigate this intractable model space, we start from a likely reference model that is suggested to match observations from previous studies \citep[e.g.,][]{Rose2022,Lam2020}. Namely, we chose a reference model of: v${}_3$ \texttt{Galaxia} model, the \model{SukhboldN20} IFMR, single lenses, and NSs (SOBHs) born with average kick velocities of 350 (100)  km s${}^{-1}$, and no PBHs. We then explore a set of perturbations (excluding adding PBHs) to this model and find the best fit model. For this best fit model we then explore adding PBHs.

To compare two models we compute the posterior probability (assuming the set of all considered models is complete) which is given by,
\begin{equation}
    p( \hyp_j|\{\DDi\} ) = \frac{\hat{p}(\{\DDi\} | \hyp_j)}{\sum_k \hat{p}(\{\DDi\} | \hyp_k)}\,,
\end{equation}
for the $j$-th model, where $\hat{p}(\{\DDi\} | \hyp_j)$ is the evidence of the model, and we have assumed equal prior probability for each model.
The model with the highest posterior probability was then used as the base-model on which the PBH residual component was appended. 

Naive evaluation of the likelihood terms in Eq. (\ref{eq:hyperpost}) is computationally intractable for two main reasons. Firstly, computation would require jointly sampling all $N_{\text{obs}}\sim10^{4}$ event posterior distributions ($\sim7\times10^{4}$ parameters) to evaluate $p(\hyp| \{\DDi\},\NOBS)$ and obtain each posterior sample for $\hyp$. Secondly, to obtain each posterior sample on $\hyp$, a $\texttt{PopSyCLE}$ simulation over each OGLE field would have to be run taking up to $\sim17,400$ core-hours.

To overcome the computational challenge of this inference problem, we make several approximations. We assume a fixed distribution of samples for each subpopulation in the $\texttt{PopSyCLE}$ simulation (and only infer the mixture fractions). We also re-use and re-weight posterior samples already obtained in \cite{Golovich2022}. Furthermore, we take advantage of the fact that some terms in the population likelihood and population efficiency can be pre-computed off-line, before inference. 
Namely, we can start with Eq.~\eqref{eq:popEventLikelihood} and substitute our population mixture model (Eq.~\eqref{eq:mixmodel}), giving
\begin{align}\nonumber
\mathcal{L}_i^{\rm obs} &= \sum_{\text{class}\in \text{classes}} \psi_{\text{class}} \int d\eparamdi p(\DDi|\eparamdi) p(\eparamdi | \text{class}) \,,\\
 &= \sum_{\text{class}\in \text{classes}} \psi_{\rm class} \mathcal{L}_{i,\text{class}}^{\rm obs} \,,
\end{align}
The integral $\mathcal{L}_{i,\text{class}}^{\rm obs}$ is tabulated for each class and each event. Similarly, $\alpha$ can be pre-computed. Starting from Eq.~\eqref{eq:alpha} (again using Eq.~\eqref{eq:mixmodel}), we obtain
\begin{align}\nonumber
\alpha &= \sum_{\text{class}\in \text{classes}} \psi_{\text{class}} \int d\D \int d \eparam  p(\trig | \D)  p(\D | \eparam) p(\eparam |\text{class}) \,\\
&= \sum_{\text{class}\in \text{classes}} \psi_{\text{class}} \alpha_{\text{class}} \,.
\end{align}
where the term $\alpha_{\text{class}}$ is tabulated and stored for re-use.

With the approximations and pre-computations, we use the \texttt{jaxns} \citep[][]{Albert2020} nested sampling algorithm implemented with \texttt{Numpyro} \citep{Phan2019} to obtain posterior samples and compute the evidence for a model. This inference takes $\approx28$ CPU core-hours ($\approx 672$ CPU core-hours) for the astrophysical (astrophysical and PBH) model.
 
\section{Results}\label{sec:results}

Due to the high model complexity in our analysis (i.e., the astrophysical systematics that must be addressed before searching for more exotic physics), the inference procedure outlined in Section~\ref{sec:framework} results in very information-rich data products.
Although in some sense a systematic effect for PBH searches, astrophysical microlensing events are also interesting in their own right, 
Therefore, the first step in our analysis is to fit an astrophysical model to OGLE's catalog of events, which is discussed in Section~\ref{sec:results.astro_model}.
Following the exploration of the astrophysical model space, the most favored astrophysical model is examined further in Section~\ref{sec:results.abundance}. 
From there, we discuss the evidence for un-modeled physics or an additional population via two, more flexible population models in Section~\ref{sec:results.residual_inference}.
Finally, we consider the specific alternative of a population of PBHs and the implications of this particular explanation in light of our results in this work in Section~\ref{sec:results.pbhpopulation}.

\subsection{Astrophysical Model Space}\label{sec:results.astro_model}

We begin by inferring which astrophysical models are favored by the data.
Our set of astrophysical models shown in Table~\ref{tab:astrophysical_models} aim to assess specific, largely uncertain physical processes encoded into each model: the Galactic model, binary sources and lenses and the physics of WD, NS  and SOBH formation.
As outlined above, we have the ability to calculate the evidence of each model fully, using the sampling algorithms and methods described above.
The evidence and posterior probability for each model is shown in Table~\ref{tab:astrophysical_model_evidence}, where we can see that the model using the slowest NS birth kicks (\model{NS Kick 150}) far outperforms the rest of the astrophysical models. Simulations of neutron star formation suggest values from 100 - 1000 km/s, with lower values associated with lower mass neutron stars \citep{Burrows2023}.
We also note that by inferring $N$ as a free parameter, our comparison of different astrophysical models does not include information about the predicted rate from each model, only its description of the population in $t_E$-$\pi_E$ space. This is a consequence of the way in which our models were constructed. As the alternative is to fix the overall rate at the predicted value (or place a restrictive prior on the parameter $N$) based on simulation, we decided to utilize the more flexible model.

\begin{figure*}
    \centering
    \includegraphics[width=\linewidth]{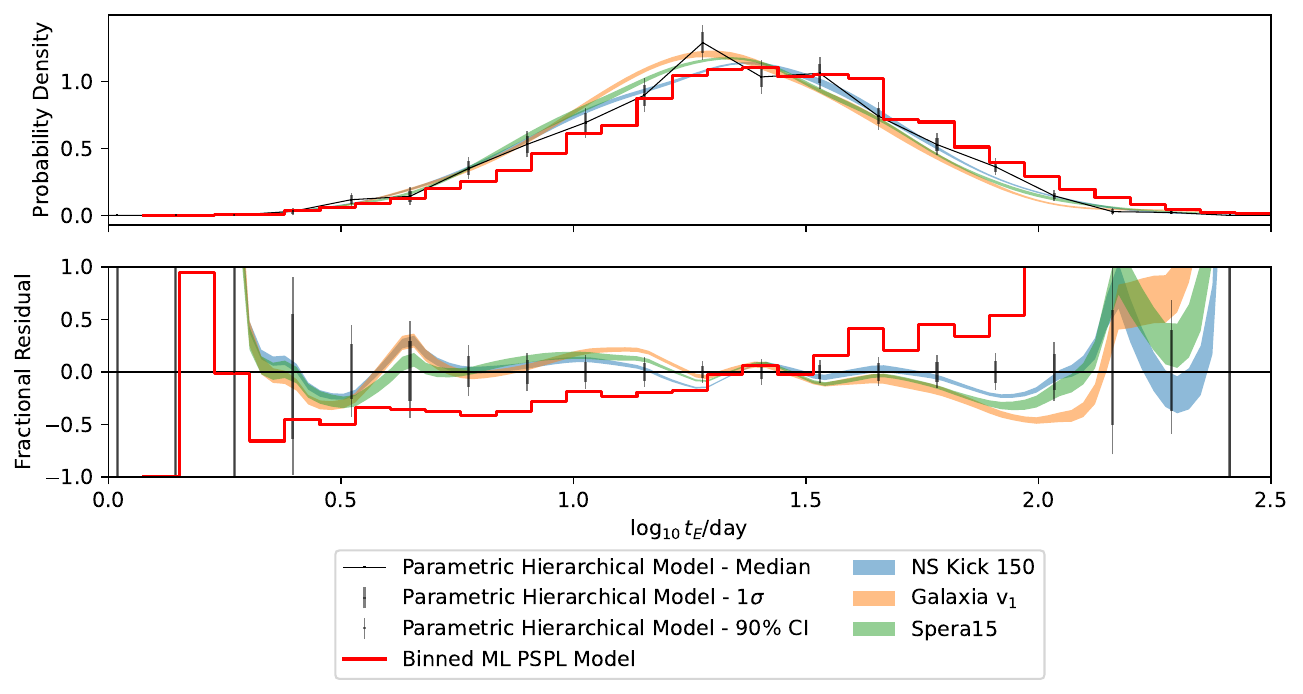}
    \caption{
    Shown above is the reconstruction of the (observed) $\log_{10} t_E$ distribution. For comparison, we show the predictions from the three best performing models (NS Kicks 150, \model{Galaxia v${}_1$} and \model{Spera15}) alongside two empirical estimates of the observed $t_E$ distribution: the binned maximum likelihood values published by OGLE~\citep[][Binned ML PSPL Model]{Mroz2019} which assumed a PSPL model in the lightcurve fitting, and the hierarchically inferred distribution using a Dirichlet model and priors defined in \citet{Golovich2022}, which used a parallax microlensing model for individual events, however fit a population distribution to the $t_E$ parameter estimates (Parametric Hierarchical Model). The median of the hyperparameter distribution is shown, with error bars reflecting the $1\sigma$ and $90\%$ confidence levels of the posterior distribution.
    The lower panel shows the fractional residual (defined in Eq.~\ref{eq:rel_diff}), comparing each model to the Parametric Hierarchical Model with the median values of its hyperparameters.
    }
    \label{fig:astrophysical_model_recon_resid}
\end{figure*}

While we fit the physics-based models using both the $t_E$ and $\pi_E$ data, we find it easier to build intuition about the model performance by examining the reconstruction of the $t_E$ distribution alone, since it is just one dimension and contains a majority of the information, due to the much better precision with which $t_E$ can be measured.
Shown in the top panel of Fig.~\ref{fig:astrophysical_model_recon_resid} is a comparison of the reconstructions of the (observed) $\log_{10} t_E$ distribution from the three best performing models (defined as having the highest posterior probabilities) and two physics-agnostic reconstructions of the distribution. The first is a histogram approximation of the distribution, using the maximum likelihood values published by OGLE~\citep{Mroz2019} (Binned ML PSPL Model). In this case, the estimation of $t_E$ for each event assumed a PSPL model, totally neglecting parallax. 
The second physics-agnostic approach is a more sophisticated representation from \citet{Golovich2022}, which uses hierarchical inference to fit a Dirichlet model to the observed distribution of inferred $t_E$ estimates (Parametric Hierarchical Model). While the hierarchical inference only considered the $t_E$ distribution, the $t_E$ estimates in this method come from fitting a parallax model to the lightcurve data, providing a more accurate, less biased estimate of the timescale. We use the same prior and model as defined in \citet{Golovich2022}.
Neither of these models provide direct insight into the \emph{physics} of the Galaxy and its populations, but serve to provide a useful metric to assess the performance of our physics-inspired and interpretable models.

In the bottom panel of Fig.~\ref{fig:astrophysical_model_recon_resid}, we show the fractional residual of each model relative to the hierarchical method used by \citet{Golovich2022}, defined as 
\begin{equation}\label{eq:rel_diff}
    \text{Fractional Residual} \equiv \frac{(p_{\text{i}}(t_E) - D (t_E))}{D(t_E)}\,,
\end{equation}
for the $i$-th model. 
The function $D(t_E)$ is Dirichlet model from the median of the parametric hierarchical inference, outlined by ~\citet{Golovich2022}.
The bands in Fig.~\ref{fig:astrophysical_model_recon_resid} are the $90$-th quantile of the residual.
The parametric hierarchical model has error bars corresponding to $1\sigma$ and 90\% confidence.

Visually, we can see from Fig.~\ref{fig:astrophysical_model_recon_resid} that all of our (much more restrictive but physically motivated) physical models are in good agreement with the parametric hierarchical model.
The best performing model (\model{NS Kick 150}) does seem to have a slight edge in reconstructing this distribution than the other two simulation-based methods shown.
It is interesting to note that both physical models outperform the estimate of the $\log_{10} t_E$ distribution from binning the ML, PSPL fit $t_E$ values.
Even with the increased complexity (yet decreased flexibility), these simulation-based models are significantly less biased than using this rough method of approximating the timescale distribution.

\begin{table}[]
    \centering
    \begin{tabular}{c|c | c}
    \hline \hline
    Model & $\ln$ Rel. Evidence & Posterior Prob.\\ \hline \hline
    \model{SukhboldN20} & -75.48 &  1.66e-33 \\
    \model{Galaxia v${}_1$} & -18.61 &  8.28e-09 \\
    \model{Galaxia v${}_2$} & -38.99 &  1.16e-17 \\
    \model{Spera15} & -21.85 &  3.24e-10 \\
    \model{Raithel18} & -34.65 &  8.96e-16 \\
    \model{NS Kick 150} & 0.00 &  1.00e+00 \\
    \model{NS Kick 200} & -67.57 &  4.53e-30 \\
    \model{NS Kick 250} & -43.48 &  1.31e-19 \\
    \model{Binaries} & -70.77 &  1.85e-31 \\
    \hline \hline
    \end{tabular}
    \caption{
    Shown in the table above are the astrophysical model evidences (relative to the best fit model, \model{NS Kick 150}) and posterior probabilities.
    The best model, by a large margin, is the \model{NS Kick 150} model, with a Bayes factor of ${\sim}10^{9}$ over the next best model (\model{Galaxia v${}_1$}).
    }
    \label{tab:astrophysical_model_evidence}
\end{table}

\begin{figure}
    \centering
    \includegraphics[width=\linewidth]{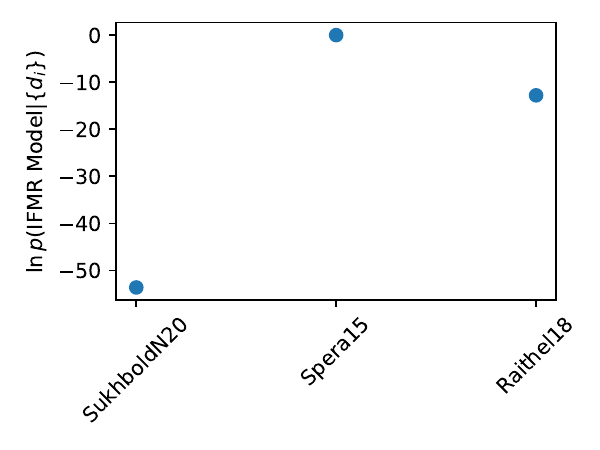}
    \caption{Shown above are the different logarithmic posterior probabilities for the different options for an IFMR model: \model{SukhboldN20}, \model{Raithel18} and \model{Spera15}.
    \model{Spera15} outperforms the other two models by a wide margin, possibly due to the more diffuse, extended mass (and subsequently $t_E$) distribution.
    }
    \label{fig:ifmr_posterior_probability}
\end{figure}

To isolate different features in the models, we can consider subsets of the model space to determine which physical theories perform the best when confronted with data.
This is a critical development: many of these subpopulations have never been interrogated at the level of population statistics compared to actual data, such as isolated stellar origin black holes.
Considering only the sets of different IFMRs, we can re-normalize the posterior probabilities to compare these three models (\model{SukhboldN20}, \model{Spera15} and \model{Raithel18}). 
These posterior probabilities are shown in Fig.~\ref{fig:ifmr_posterior_probability}.
\model{Spera15} outperforms both of the other models by a wide margin. 
This improvement in modeling could be due in part to the long tail in the mass distribution that \model{Spera15} encodes (in contrast to \model{Raithel18} and \model{SukhboldN20}). 
\citet{Rose2022} examined implications of these different IFMR models, showing that \model{Spera15} indeed leads to a longer tail in the $t_E$ distribution, as expected from a wider mass distribution.
Furthermore, the NS mass distribution is wider in the \model{Spera15} model than in \model{Raithel18} and \model{SukhboldN20}.
This leads to a wider $t_E$ distribution for the NS lenses, in conjunction with the wider distribution of $t_E$ for the SOBH.
The pair of these two modeling implications could explain the discrepancy in the modeling capability of \model{Spera15}, compared to \model{Raithel18} and \model{SukhboldN20}: there is a preference for smoother, wider representations of the observed $t_E$ distribution.

\begin{figure}
    \centering
    \includegraphics[width=\linewidth]{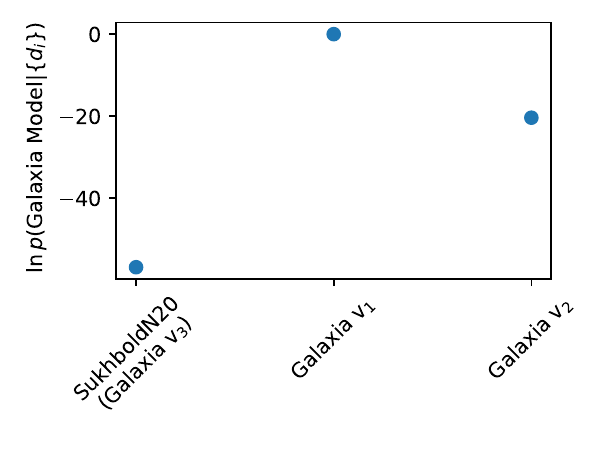}
    \caption{
    Shown above are comparisons of the different Galactic Bulge models used in this study (see Table~\ref{tab:galaxia_models}).
    \model{Galaxia v${}_1$} and \model{Galaxia v${}_2$} outperform \model{Galaxia v${}_3$}, suggesting the average velocity distribution in the Milky Way and not the spatial distribution of objects drives the effectiveness of microlensing surveys to probe these differences.
    }.
    \label{fig:galaxia_posterior_probability}
\end{figure}

In addition to investigating the IFMR models, we can also consider the general Galactic Bulge spatial orientation and average dynamics.
We have three separate models of the Galactic Bulge, denoted \model{Galaxia v${}_1$}, \model{Galaxia v${}_2$} and \model{Galaxia v${}_3$} (referred to as \model{SukhboldN20} in this work, as \model{Galaxia v${}_3$} is our assumed galactic model). 
These different models are discussed in detail in \citet{Lam2020}, but generally differ in the angle of the Galactic bar, size of the bar and the pattern speed of the galaxy.
While not appreciably changing the mass distribution of lenses in the galaxy, it does change the average spatial and dynamical relationship between sources and lenses, altering the distributions in $t_E$ and $\pi_E$ through these changes in auxiliary distributions.
The posterior probability between these models is shown in Fig.~\ref{fig:galaxia_posterior_probability}.
Both \model{Galaxia v${}_1$} and \model{Galaxia v${}_2$} outperform \model{Galaxia v${}_3$}. 
This observation immediately isolates the velocity dispersion relationship and the bar patternspeed as being highly influential over the distributions of microlensing observables. 
\model{Galaxia v${}_1$} and \model{Galaxia v${}_2$} both differ from \model{Galaxia v${}_3$} in their kinematics.
However, \model{Galaxia v${}_1$} also differs from \model{Galaxia v${}_3$} in its spatial distribution, unlike \model{Galaxia v${}_2$}.
As \model{Galaxia v${}_1$} also outperforms \model{Galaxia v${}_2$}, this analysis suggests that while the dominant effect is the kinematics of the galaxy, the spatial distribution also has a secondary effect on the analysis.
This should be unsurprising, as microlensing surveys are dominantly sensitive to the distribution of characteristic times, over the distribution of $\pi_E$ (which is a much harder parameter to infer for most events).
Again, we note that the structure of these models encode $N$ as a free parameter, as we sought to err on the side of flexibility for our astrophysical models. Past work has found differing results, but primarily relied on rate information for their analysis~\citep{Lam2022}.
The distribution of times is intimately connected to the relative velocity of the lens and source, directly connected to the average speed of objects in the Milky Way.
Meanwhile, the difference in bar spatial configuration seems to be a subtle change to the spatial distribution of lenses and sources, as \model{Galaxia v${}_1$} performs comparably to \model{Galaxia v${}_2$}.

\subsection{Astrophysical Abundances and Rates}\label{sec:results.abundance}

\begin{figure}
    \centering
    \includegraphics[width=\linewidth]{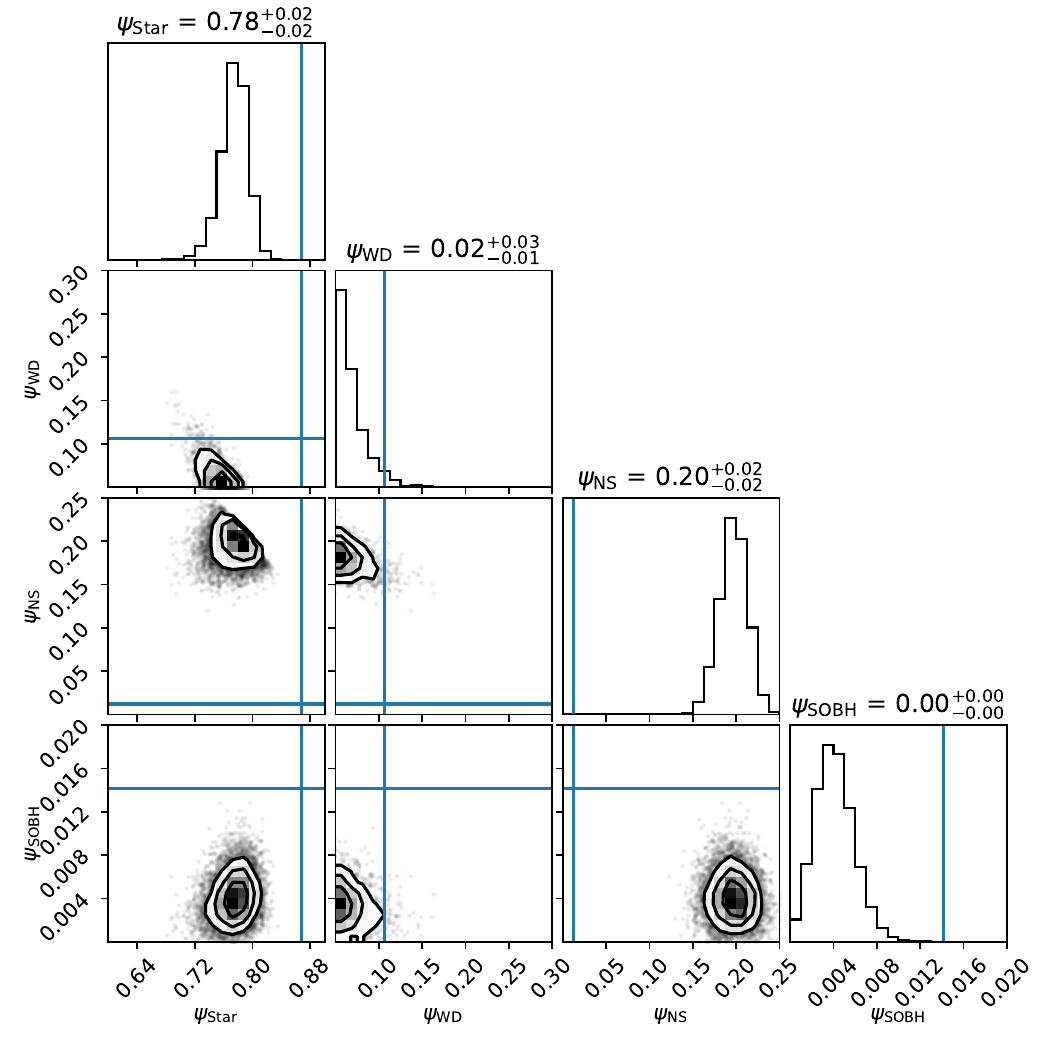}
    \caption{
    The posterior distributions for the mixture fractions are shown above, with the expected fractions from simulation shown in blue. 
    The NS population seems to be the biggest outlier, when compared with simulation, with compensating shifts in the Stellar and SOBH  populations.
    The WD population is consistent with simulation, albeit with wider posteriors. This is unsurprising, as the WD population falls very close to the stellar and NS populations in the space of $t_E$-$\pi_E$, more so than the NS population.
    }
    \label{fig:favored_astro_model_mixture_fractions}
\end{figure}

\begin{figure}
    \centering
    \includegraphics[width=\linewidth]{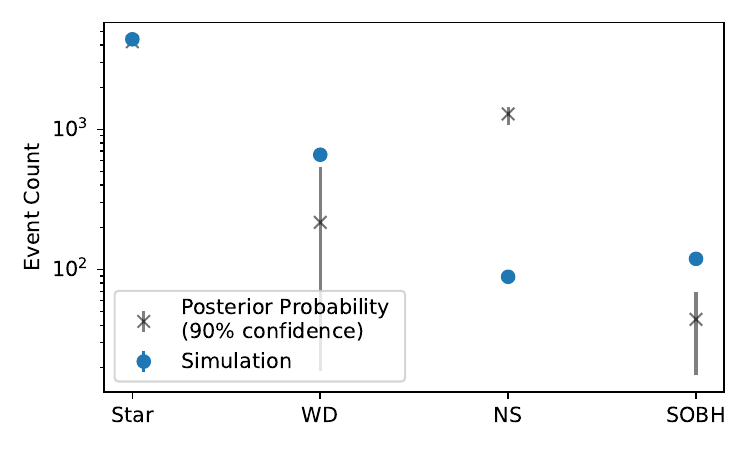}
    \caption{
    A comparison of simulation (blue points, error bars reflect expected Poisson sampling errors) to the posterior probability (black markers, error bars show 90\% confidence) for the event rate, per subpopulation. 
    The estimates above are predictions for the expected \emph{detected} events, i.e., taking into account OGLE's detection efficiency curves, survey duration and survey area.
    The agreement here is subtly different than the mixture fraction posteriors, as this encodes additional information about the overall rate (including correlations with the overall rate and the rates for each subpopulation).
    }
    \label{fig:favored_astro_model_rates}
\end{figure}

With an overall astrophysical model selected via each model's posterior probability, we can examine the best-fitting model in more detail. 
The parameters of the most interest are the relative abundances (divorced from the uncertainty of rate calculations) and the expected detection rates of the various subpopulations.
The posterior distribution for the mixture fractions are shown in Fig.~\ref{fig:favored_astro_model_mixture_fractions}, while the expected detection rate is shown in Fig.~\ref{fig:favored_astro_model_rates}.

The inferred mixture fractions of the different populations show a slight shift towards centrally located populations compared to the simulated expectations, both in mass and timescale. 
The populations at the extreme ends of the distribution (Stars and SOBH subpopulations) are both inferred to be systematically lower than expected by simulation, as is the WD population.
However, the NS subpopulation is so high that we infer they make up a larger fraction of the lensing population that WD do.

This over-abundance of NSs is pronounced, but we caution taking this result at face value. Our models in Table.~\ref{tab:astrophysical_models}, while improving and extending past modeling efforts in many novel ways, do not capture systematics not explicitly outlined in the description of said models. Critical assumptions related to the \emph{internal} structure and dynamics of these different populations (like the stellar mass distribution or spatial profile, for example) were not assessed. This inferred over abundance could be symptomatic of these biases not explored in the current work. The sections that follow aim to further disentangle and contextualize these results by using an even more flexible model to perform the inference. 

\subsection{Inference with a Residual subpopulation component}\label{sec:results.residual_inference}

Now that a best-performing astrophysical model has been identified (the \model{NS Kick 150} model), we can consider how we might investigate systematics beyond the standard physics encoded into \texttt{PopSyCLE}. 
To do that, we will include our PBH and astrophysical residual models (parameterized by mass, as outlined in Sec.~\ref{sec:mass_distribution_modeling}) into the total model, now denoted as \model{NS Kick 150 + PBH} and \model{NS Kick 150 + Astro. Residual}, respectively.

Following the methods applied to the suite of astrophysical models, we can first compare the Bayes' factor between the \model{NS Kick 150}, \model{NS Kick 150 + PBH}  and \model{NS Kick 150 + Astro. Residual} models. 
With the priors specified in Table~\ref{tab:flexibleHyperparameters}, this comes out to be $\ln\mathcal{B}^{\text{\model{NS Kick 150 + PBH}}}_{\text{\model{NS Kick 150}}} =20.23$ and $\ln\mathcal{B}^{\text{\model{NS Kick 150 + Astro. Residual}}}_{\text{\model{NS Kick 150}}} = 2.53$. 
This reflects a large preference for the PBH model but only a mild preference for the astrophysical residual model.
The mild preference for the astrophysical residual model over the standard astrophysical model gives evidence for possible systematics in the mass distribution of lenses, but the much larger evidence for the PBH residual model suggests the need for refined modeling going beyond the mass distribution.
As the mass distributions are similarly modeled in both residual models, it would seem there is a boost in the PBH model's ability to explain the data due to differing assumptions about the distribution of lenses and velocities in the galaxy.

\begin{figure}
    \centering
    \includegraphics[width=\linewidth]{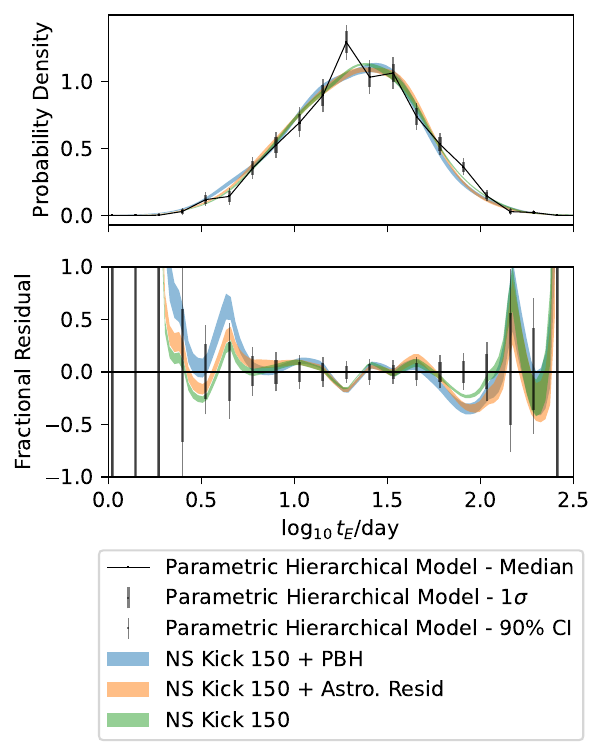}
    \caption{
    Shown above is a comparison of the total $t_E$ distribution (posterior predictive, 90\% CI) between the \model{NS Kick 150 + PBH} model, \model{NS Kick 150 + Astro. Residual} model and the \model{NS Kick 150} model.
    While there is a minor shift of the overall width of the distribution, the discrepancy between the different models, especially in the case of the \model{NS Kick 150 + PBH} model, is much smaller than that between different astrophysical models (Fig.~\ref{fig:astrophysical_model_recon_resid}).
    }
    \label{fig:pbh_astro_tE_comp}
\end{figure}

\begin{figure}
    \centering
    \includegraphics[width=\linewidth]{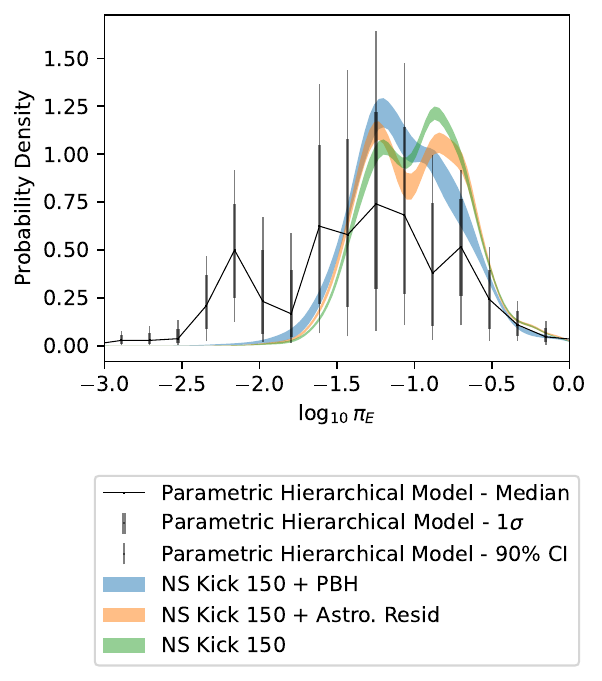}
    \caption{
    Shown above is a comparison of the total $\pi_E$ distribution (posterior predictive, 90\% CI) between the \model{NS Kick 150 + PBH} model, \model{NS Kick 150 + Astro. Residual} model and the \model{NS Kick 150} model.
    In this case, with much less information being contained in each individual event's posterior for $\pi_E$, the different models have significantly different predictions, giving a possible hint as to how the \model{NS Kick + PBH} model can outperform both the \model{NS Kick 150 + Astro. Residual}  and \model{NS Kick 150} models, despite having similar predictions for the $t_E$ distribution.
    }
    \label{fig:pbh_astro_piE_comp}
\end{figure}

To better understand this discrepancy between the two residual models, we can look at some of the key posterior predictive distributions for quantities of interest, beginning with the $t_E$ and $\pi_E$ distributions.
These are shown in Fig.~\ref{fig:pbh_astro_tE_comp} and Fig.~\ref{fig:pbh_astro_piE_comp}, respectively.
The ``parametric hierarchical model'' from \citet{Golovich2022} was applied to the observed $\pi_E$ distribution, with 30 bins between $\pi_E \in [10^{-5}, 3] $ spaced logarithmically. 
The prior to infer the observed $\pi_E$ distribution was a uniform Dirichlet, i.e., a concentration vector $\boldsymbol{a} = \boldsymbol{1}$.
First examining Fig.~\ref{fig:pbh_astro_tE_comp}, we see there is general agreement between the three models, with the \model{NS Kick 150 + PBH} showing a slight preference for a (slightly) broader $t_E$ distribution.
Meanwhile, the discrepancy is substantially larger in the case of $\pi_E$, showing both residual models move towards a lower distribution on $\pi_E$.
The combination of these distributions (taken as a pair of one dimensional posterior predictive distributions alone, a flawed representation of the situation) suggests the $t_E$ modeling of these different models has similar performance, while the data prefers models that can predict lower distributions in $\pi_E$.
The discrepancy in $\pi_E$ alone is a first hint at why the residual models have better explanation power. A better understanding of the situation comes when we consider the joint space of $t_E$-$\pi_E$. Even further insight comes from considering the physical implications for these preferences in parameter space. Both of these next steps are considered below.

\begin{figure}
    \centering
    \includegraphics[width=\linewidth]{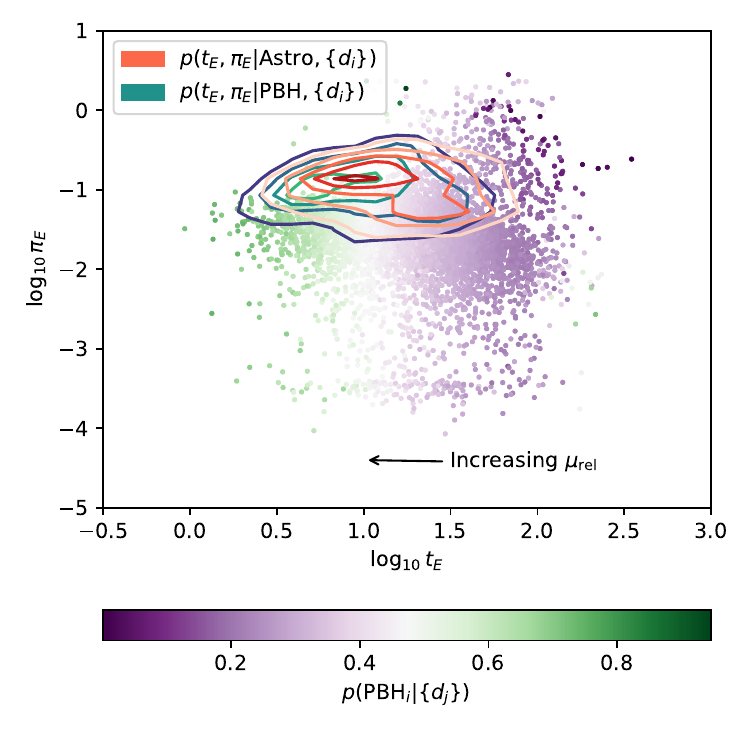}
    \caption{
    Medians of the likelihoods in $t_E$ - $\pi_E$ for each OGLE-IV event analyzed by \citet{Golovich2022} (calculated by finding the median value of the posterior, weighted by reciprocal of the prior).
    Each median point is colored according to its probability of being a PBH for this model (averaged over the posterior of the \model{NS Kick 150 + PBH} model), which serves as a proxy for each event's contribution to the evidence for needing a PBH component to the model. 
    The predictions for the PBH component (only) of the \model{NS Kick 150 + PBH} model is shown above, taking the median values of each mixture component of the mass distribution, $\boldsymbol{\beta}$.  
    For comparison, the prediction for the \model{NS Kick 150} model (including all astrophysical components) is shown in red, again derived from the median of the posterior distribution for each population component.
    The contours for both models are uniformly spaced between $0$ and $1.5$. 
    The arrow in the top panel indicates the shift in parameters from increasing $\mu_{\text{rel}}$ (and only $\mu_{\text{rel}}$).
    From the figure, it appears that better support for events at low $\pi_E$ and low $t_E$ is helping the \model{NS Kick 150 + PBH} outperform its astrophysical counterpart.
    } \label{fig:pbh_astro_comparison_individual_events}
\end{figure}

While these one dimensional posterior predictive distributions can begin to hint at the underlying reason for the preference for the PBH residual model, we must next look at the full covariant space of $t_E$ and $\pi_E$.
The distributions expected from the PBH residual model (separated by bin) and the \model{NS Kick 150} model predictions are shown in Fig.~\ref{fig:pbh_astro_comparison_individual_events}, compared to point estimates (medians) of the likelihood distributions for each of the individual events.
Each of the median values are colored according to their probability of involving a PBH lens, averaged over the posterior distribution of the \model{NS Kick 150 + PBH} model.
Mathematically, this is defined as 
\begin{align}\nonumber
    p(\text{PBH}_i | \{d_i\}) &\propto \int p(d_i | t_E, \pi_E ) \\ \nonumber
    &\times p(t_E, \pi_E | \text{PBH}, \boldsymbol{\beta}_{\text{PBH}}, \boldsymbol{\psi}, N)\\\nonumber
    &\times p(\text{PBH} | \boldsymbol{\beta}_{\text{PBH}}, \boldsymbol{\psi}, N) \\ \nonumber
    &\times p(\boldsymbol{\beta}_{\text{PBH}}, \boldsymbol{\psi}, N | \{d_i\}) \\
    &\times d \boldsymbol{\beta}_{\text{PBH}} d\boldsymbol{\psi} d N d\pi_E d t_E\,.
\end{align}
In summary, we are asking what the probability is that the $i$-th event involves a PBH lens ($\text{PBH}_i$), given we have observed a set of data $\{d_i\}$ for $i\in \{1 \ldots N_{\text obs}\}$.
The overall normalization comes from assuming the the probabilities that a lens belongs to each class $\{ \text{Star}, \text{WD},\text{NS},\text{SOBH},\text{PBH}\}$ sum to one.
This quantity is the posterior probability that the $i$-th event involves a PBH lens (which is not of immediate concern in this work), but we can also use it as a metric to determine each event's contribution to the PBH model's evidence.
The events with the highest probability of being a PBH are the same events which drive the preference for a PBH subpopulation component in the model.

Fig.~\ref{fig:pbh_astro_comparison_individual_events} gives further guidance as to why the \model{NS Kick 150 + PBH} model outperforms the rest of the possible models, even the \model{NS Kick 150 + Astro. Residual}. 
From Fig.~\ref{fig:pbh_astro_tE_comp} and Fig.~\ref{fig:pbh_astro_piE_comp}, we can see that the model which can maintain or slightly improve the fit in $t_E$, but can reach to much lower average $\pi_E$ values is the favored model.
In the covariant space, this means favoring models which have better population coverage at the low $t_E$ and low $\pi_E$ quadrant of the space, around $\log_{10} t_E \sim 0.5$ and $\log_{10} \pi_E \sim -1.5$.

\begin{figure}
    \centering
    \includegraphics[width=\linewidth]{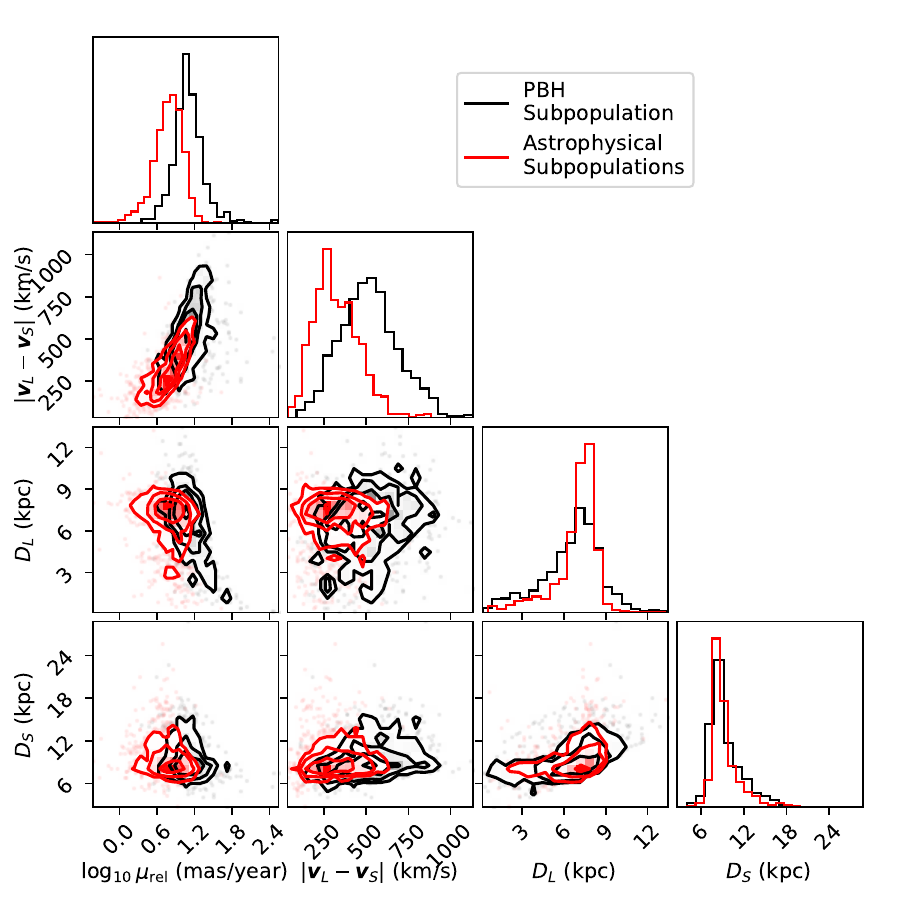}
    \caption{
    A comparison of the auxiliary parameters $\{\mu_{\text{rel}}, |\boldsymbol{v}_L - \boldsymbol{v}_S|, D_L, D_S\}$ for the PBH population (shown in black) assuming a log-uniform mass distribution and the astrophysical population (shown in red).
    For PBH lenses, they indeed follow a modified distribution in their distances from Earth, but more importantly in the current context, also follow a different velocity distribution than the astrophysical microlensing sources in the galaxy.
    This difference causes a shift in the relative angular velocity $\mu_{\text{rel}}$ and relative velocity $|\boldsymbol{v}_L - \boldsymbol{v}_S|$.
    }
    \label{fig:auxiliary_parameters}
\end{figure}

What, then, physically moves the PBH subpopulation to the lower $\pi_E$ distributions in this part of parameter space?
In calculating the $t_E$ and $\pi_E$ distributions, we need more information than just the mass of the lens: from Eq.~\ref{eq:tEDef} and Eq.~\ref{eq:pi_E}, we need to implicitly marginalize over distributions for the distances and relative velocity, both of which are fundamentally different for the PBH subpopulation as it follows the DM Halo profile (discussed in Sec.~\ref{sec:forwardModeling}).
As both the \model{NS Kick 150 + Astro. Residual} and \model{NS Kick 150 + PBH} models have a flexible mass distribution, the key difference in their ability to describe the data lies in these auxiliary distributions. 
Given we have simulation output which retains all of this information, we can compare the predicted distributions for these auxiliary parameters between the PBH subpopulation and the astrophysical subpopulations.
This three dimensional distribution for both the PBH subpopulation and the astrophyscial population using the simulation underlying the \model{NS Kick 150 + PBH} model is shown in Fig.~\ref{fig:auxiliary_parameters}.

From Fig.~\ref{fig:auxiliary_parameters}, we see that the dimension with the largest change (directly related to microlensing parameters) is the relative angular velocity parameter, $\mu_{\text{rel}}$.
When considering how the PBH population differs from the astrophysical ones, this difference is key: with the  allowed astrophysical model space, the distributions of $\mu_{\text{rel}}$ changes very little. 
Even when adjusting the pattern speed with \model{Galaxia v${}_1$} or \model{Galaxia v${}_2$}, the sources and lenses are generally drawn from the same velocity distributions producing similar distributions of \emph{relative} velocities. 
There is very little flexibility to impact the \emph{relative} velocity distribution.
The PBH subpopulation is entirely different, as it is one of the only lens populations we have considered that has a velocity distribution fundamentally different to and independent from the source distribution. 
This can be seen both in the $\mu_{\text{rel}}$ distribution, but also in the relative velocity distribution, $|\boldsymbol{v}_L - \boldsymbol{v}_S|$.
We focus on the former, as it more directly ties with the microlensing parameters of interest, but the difference in velocities ties more fundamentally into the physical process sourcing this discrepancy.
The NS and SOBH populations are similar in that they are drawn from \emph{modified} velocity distributions (relative to the stellar population) because of their natal kicks, but are still fundamentally related to the stellar velocity distribution.
Ultimately, this effect gives the PBH subpopulation a unique ability to describe the data, not well mimicked by other astrophysical populations.
In contrast to flexibility in mass distribution, a fundamentally different relative velocity distribution opens up an orthogonal direction to model in the space, one that shifts distributions in a single direction ($t_E$) as opposed to jointly shifting distributions in $t_E$ and $\pi_E$.

\subsection{Inference on the PBH Population}\label{sec:results.pbhpopulation}

\begin{figure}
    \centering
    \includegraphics[width=\linewidth]{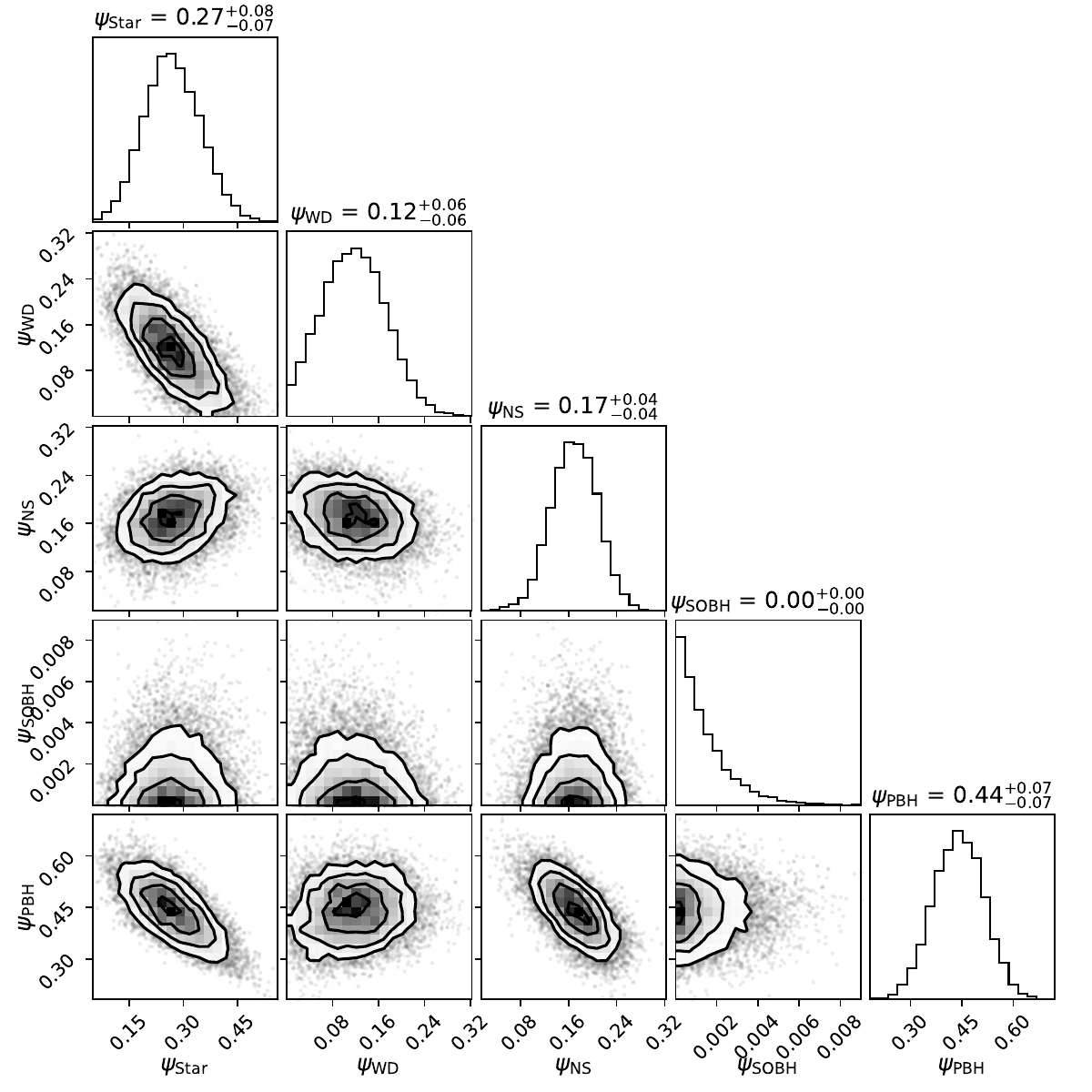}
    \caption{
    Shown above is the posterior distribution on the mixture components for the \model{NS Kick 150 + PBH} model. 
    The PBH model takes up a substantial fraction of the lens population (${\sim}50\%$), mostly at the cost of lowering the number of Stars and SOBHs.
    The WD and NS populations are not substantially modified by the introduction of the PBH population.
    }
    \label{fig:pbh_mixture_components}
\end{figure}

Considering the model selection results of the previous section, we can ask what the implications are for the hypothetical population of PBHs.
The first quantity of interest is the mixture fractions of the population model, $\boldsymbol{\psi}$, which reflect the fraction of the lensing population associated with each class of object.
This is shown in Fig.~\ref{fig:pbh_mixture_components}.
The PBH population makes up a substantial number of the lenses in this population model (${\sim}50\%$).
This shift to include more PBH lenses comes at the cost of lowering the number of Stellar and SOBH lenses, both lowering by approximately ${\sim}75\%$ of their original fractions. 
The mixture fraction posterior reflects the improvement of the \model{NS Kick 150 + PBH} model over the \model{NS Kick 150} model, as the PBHs contribute to the model's ability to explain the data in a meaningful way, so a posterior consistent with no PBH lenses would be inconsistent with the Bayes factor analysis.

\begin{figure}
    \centering
    \includegraphics[width=\linewidth]{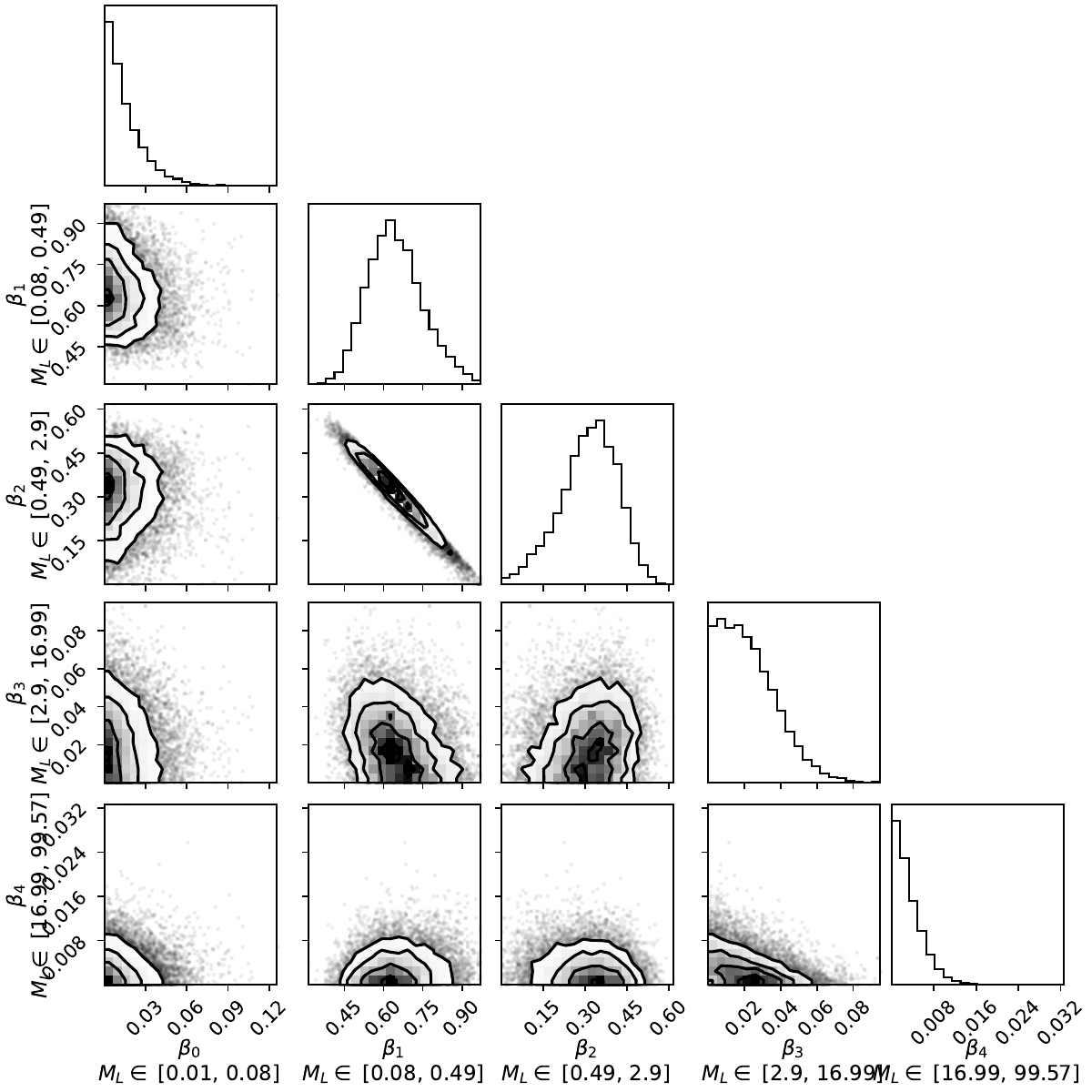}
    \caption{
    Shown above is the posterior distribution on the mixture components for the PBH mass distribution ($\boldsymbol{\beta}$).
    The lowest and two highest bins are consistent with zero, suggesting these ranges of PBH mass are not well supported or needed by the data.
    Meanwhile, $\beta_1$ and $\beta_2$ are inconsistent with zero, reflecting that this mass range ($M_{\text{PBH}} \sim 0.08 - 2.9 M_{\odot}$) contributes meaningfully to explaining the data.
    }
    \label{fig:pbh_mass_bins_posterior}
\end{figure}

\begin{figure}
    \centering
    \includegraphics[width=\linewidth]{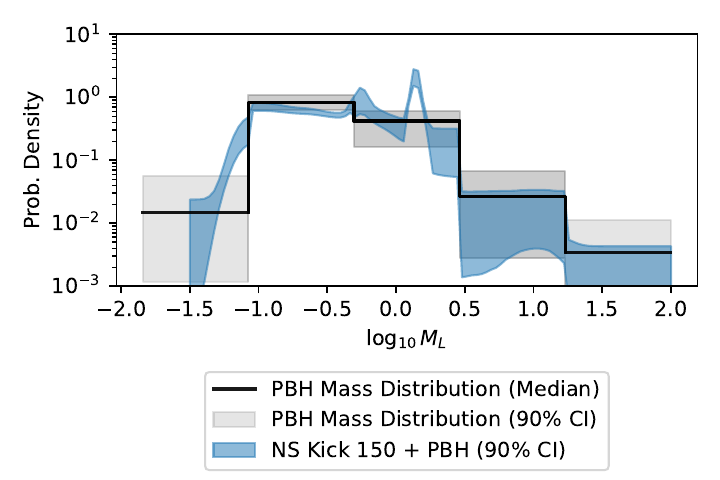}
    \caption{
    The above figure shows the PBH mass distribution (black shaded) and total lens mass distribution (blue) for the \model{NS Kick 150 + PBH} model.
    We can see that there is strong degeneracy in the modeling of the PBH mass spectrum and the total lens mass distribution due to lens mass roughly scaling with the square root of the characteristic microlensing time.
    However, as discussed in previous sections, this degeneracy in the mass spectrum is only a partial degeneracy, as PBH lenses of a given mass have different $t_E$-$\pi_E$ predictions than astrophysical lenses of the same mass.
    }
    \label{fig:pbh_mass_fn_posterior}
\end{figure}

To further investigate what parts of the PBH model are helping to better explain the data, we can next look at the inferred mass distribution of the PBH subpopulation.
This is visualized in two ways: once via showing the posterior distribution on the mixture components of the PBH mass spectrum ($\boldsymbol{\beta}$; Fig.~\ref{fig:pbh_mass_bins_posterior}) and once in terms of the actual posterior predictive distribution for the PBH mass spectrum ($p(\log_{10} M_L)$; Fig.~\ref{fig:pbh_mass_fn_posterior}).
Fig.~\ref{fig:pbh_mass_bins_posterior} shows the majority of events come from two mass bins in the middle of the distribution, between $0.08-2.9M_{\odot}$, which directly overlaps with the astrophysical population (particularly dominant in the regime populated by Stars). 
This range of PBHs is consistent with the shift in the relative abundance of stellar lenses in the population model, as the overall stellar makeup of the lens population dropped significantly in favor of PBH lenses.
Again, however, it is important to mention that just because these PBH lenses have similar masses to the Stellar lens population, their distribution in $t_E$-$\pi_E$ is shifted because of these auxiliary parameter distributions (like the relative angular velocity).

Additionally, the PBH subpopulation is consistent with zero at the lowest mass bin (sub-stellar), and above ${\sim}2 M_{\odot}$.
The former observation appears to suggest that the low timescale (${\lesssim}1$ day), low mass regime is well modeled by \texttt{PopSyCLE}. 
Although, this must come with a few caveats. 
First, this timescale is poorly probed by OGLE's low-cadence observations, and so the inference is mostly driven by non- or low-detection rate of events in this $t_E$-$\pi_E$ regime and extrapolated via OGLE's assessment of the survey's detection efficiency.
Secondly, working in the covariant space means that we are restricted in saying that low mass \emph{PBH} lenses are not favored by this study, but other low mass objects (e.g., FFPs and BDs, which are not included in \texttt{PopSyCLE}) would need careful treatment to better assess their consistency with the data.
Similarly in the high mass, long timescale regime, the consistency of the PBH mass bins with zero suggests a general agreement of \texttt{PopSyCLE} with OGLE's data, albeit with the same caveats that the low rate of detections makes this conclusion heavily conditioned on OGLE's assessment of their detection efficiency.
As pointed out by \citet{Mroz2019}, the high timescale detection efficiency curve is susceptible to bias, as parallax and binary systems were not taken into account in the calculation of the detection efficiency, both of which will have larger impacts at long timescales as compared to short timescales.

Meanwhile, Fig.~\ref{fig:pbh_mass_fn_posterior} shows the comparison of the different posterior predictive distributions for the mass of PBHs and the total lens mass distribution.
From this figure, we can see that there is an expected degeneracy in the PBH mass distribution and the total lens mass distribution.
This is no surprise, as $t_E$ still generally scales with the (square root) of the lens mass even though there are nuanced differences in the exact shape and location of the distributions in $t_E$-$\pi_E$ for the different subpopulations (discussed above).   
While the degeneracy could be minimized with more restrictive prior densities for the mixture components $\boldsymbol{\psi}$, the prior would have to be fairly restrictive to force the PBH mixture component to be consistent with zero (reflective of the large Bayes factor in favor of the PBH model over \model{NS Kick 150}).
The result that the total mass distributions follow a similar form suggests future efforts should focus on the spatial and velocity distribution profiles of the galaxy, as was pointed out earlier in this section. 

\begin{figure}
    \centering
    \includegraphics[width=\linewidth]{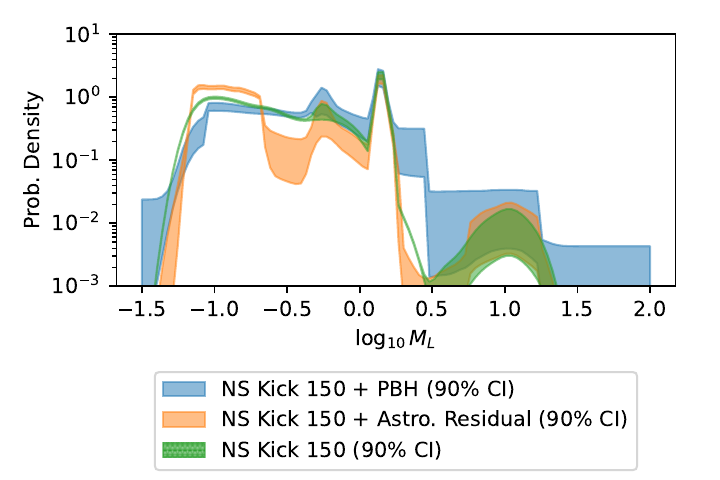}
    \caption{
    Shown above is a comparison of the different total lens mass distributions: \model{NS Kick 150 + PBH} (blue), \model{NS Kick 150 + Astro. Residual} (orange) and \model{NS Kick 150} (green).
    While there is general agreement of the different mass functions, we have seen in previous sections that this does not necessarily correspond to similar distributions in $t_E$ and $\pi_E$.
    Focusing on the finer details, the PBH model prefers a flatter mass distribution at low masses, but fails to replicate the features caused by the sharp mass distributions of WDs and NSs (the peaks at $\log_{10} M_L \sim-0.25$ and $\log_{10} M_L \sim0.2$, respectively).
    The \model{NS Kick 150 + Astro. Residual} model prefers a sharper mass function, opposite the posterior predictive distribution of the PBH model.
    This result underscores the importance of understanding the proper calculation of the $t_E$ and $\pi_E$ distributions, as they produce similar timescale distributions shown in Fig.~\ref{fig:pbh_astro_tE_comp} (albeit, with different $\pi_E$ distributions shown in Fig.~\ref{fig:pbh_astro_piE_comp}).
    }
    \label{fig:pbh_mass_fn_posterior_comparison}
\end{figure}

A comparison of the different total lens mass distributions from the three models is shown in Fig.~\ref{fig:pbh_mass_fn_posterior_comparison}.
There is general agreement in the different models, but there are also small deviations as well.
While the \model{NS Kick 150 + Astro. Residual} favors a sharper mass distribution in the Stellar regime ($\lesssim 0.5 M_{\odot}$), the \model{NS Kicks 150 + PBH} favors a wider distribution in lens mass at low masses. 
Furthermore, we can see that the PBH component cannot compensate for the effects of the narrow distributions of the WD and NS lenses (at approximately $0.5M_{\odot}$ and $1.5M_{\odot}$, respectively), as those are well represented in all models.
Finally, at high mass we see a strong degeneracy with the SOBH population, as there is little discriminating power at these longer timescales with low event rates.

Finally, a conservative reader might choose to use this analysis to place upper limits on the size of a hypothetical population of PBHs. 
To translate constraints on the lens population mixture fraction $\psi_{\text{PBH}}$ to $f_{\text{DM}}$ takes a few extra steps, however.
The mixture fraction we infer represents a sub-selection of PBHs chosen according to satisfy the definitions used by OGLE, namely that they form a microlensing system with a source star with a magnitude of $I < 21$ and a normalized impact parameter $|u_0| < 1$. 
To account for this selection effect, we must use the simulation output, which is a numerical estimate of the fraction of events which would be expected to persist beyond this cut.
We first calculate the total mass in PBHs contained in the simulation output corresponding to a universe with $f_{\text{DM}} = 1 $ via 
\begin{equation}
    M_{\text{PBH,simulation}} = \frac{\hat{A} \langle M \rangle_\text{simulation} N_{\text{PBH,simulation}}}{f_{\text{DM,simulation}}}\,,
\end{equation}
where $\hat{A} $ is a scaling factor to translate between our simulated field area to OGLE's field area, equal to $112\times 1.4 /( 18 \times 0.3) $\footnote{As a reminder, we only consider the low cadence fields from OGLE-IV.}.
$\langle M\rangle_\text{simulation}$ is the average mass of a PBH lens in the simulation, and $N_{\text{PBH,simulation}}$ is the number of PBH lenses which survive the cuts outlined above.
Finally, the quantity $f_{\text{DM,simulation}} = 10$ is the $f_{\text{DM}}$ used in the simulation, which we increased to ten to decrease numerical noise.

From here, we then calculate the average mass of a PBH lens (for each sample of the posterior)
\begin{align}\nonumber
    \langle M \rangle &= \int M p(\log_{10} M) d\log_{10} M\,,\\
    & = \sum_i \frac{\beta_i N_B}{\Delta \log_{10} M } \frac{1}{\ln{10}} (M_{i+1} - M_{i} ) \,,
\end{align}
where we can analytically compute the average because of the form of the mass distribution in Eq.~\ref{eq:pbh_mass_distribution}.
The quantities $M_i$ and $M_{i+1}$ represents the lens masses at the boundaries of the $i$-th bin edge.
With our new estimate of the average mass of PBHs in our model defined by a posterior sample, we can calculate the corresponding $f_{\text{DM}}$ needed to produce that processed subpopulation of PBHs.
\begin{equation}
    f_{\text{DM,posterior}} = \frac{N \psi_{\text{PBH}}\langle M \rangle}{M_{\text{PBH,simulation}}}\,,
\end{equation}
where $N$, $\psi_{\text{PBH}}$ and $\langle M \rangle$ are derived from a single draw from the hyper-parameter posterior distribution.
Ultimately, this operation simply finds the ratio of the PBH mass contained in the OGLE survey footprint and divides it by the expected mass of PBHs (for a universe with $f_{\text{DM}} = 1$) derived from simulation.

\begin{figure}
    \centering
    \includegraphics[width=\linewidth]{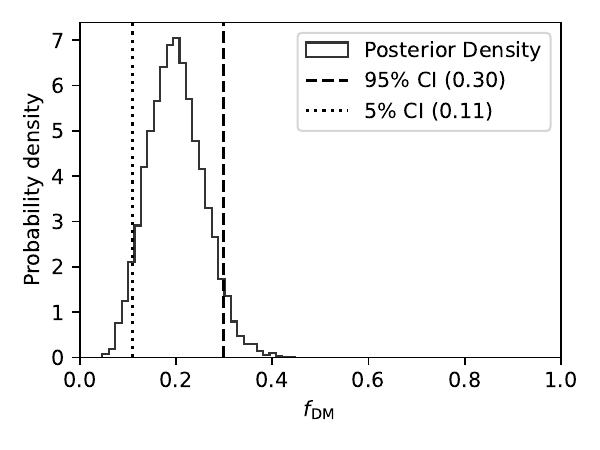}
    \caption{
    The posterior distribution on $f_{\text{DM}}$ is shown above, inferred using the \model{NS Kick 150 + PBH} model.
    This posterior distribution is marginalized over the mass distribution of the PBH lensing events, as well as the other astrophysical modeling.
    We also show the $95\%$ ($5\%$) (one-sided) CI for the posterior, inferred to be 0.30 (0.11).
    }
    \label{fig:fdm_posterior_constraint}
\end{figure}

We show this analysis, marginalized over the PBH mass distribution, in Fig.~\ref{fig:fdm_posterior_constraint} with both the $95\%$ and $5\%$ (one-sided) CI annotated.
With the models described in this work, we estimate an upper limit (at 95\% CI) on $f_{\text{DM}}$ of $0.30$.
This constraint is well above past work (summarized in Fig.~\ref{fig:current constraints}), but the analysis in this work expands on those past results by considering extended mass distributions (also being done in the context of GWs~\cite{Franciolini2021}, for example) and examined a set of data with substantially higher numbers of microlensing events, compared to past work in the microlensing field. 

\begin{figure}
    \centering
    \includegraphics[width=\linewidth]{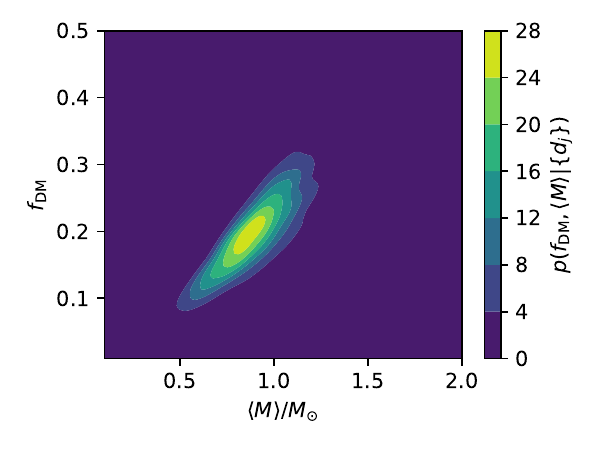}
    \caption{
    Shown above is the posterior probability of the average mass the PBH mass spectrum and $f_{\text{DM}}$.
    For clarity, let us reiterate that this is the posterior distribution on $f_{\text{DM}}$-$\langle M \rangle$, and therefore the 95\% contour and \emph{out} corresponds to the constraint.
    This figure is more closely aligned with typical figures of the constraints (see Fig.~\ref{fig:current constraints}), albeit our depiction is more nuanced.
    We have marginalized over the exact shape of the distribution, and are reducing these results down to the first moment of the PBH mass distribution, as opposed to using a monochromatic mass function as the rest of the constraints in Fig.~\ref{fig:current constraints} do.
    The OGLE data, coupled with our analysis and models and taken at face value, suggests a $f_{\text{DM}}\sim 0.2$ and $\langle M\rangle \sim 1$ to be the preferred.
    }
    \label{fig:fdm_average_mass}
\end{figure}

This constraint is also shown in the $\langle M \rangle$-$f_{\text{DM}}$ space, in Fig.~\ref{fig:fdm_average_mass}.
Here, we get a better picture of the correlation between the DM fraction and the shape of the PBH mass distribution, albeit in a reduced form as the average mass is only the first moment of the mass distribution. 
From this figure, we see that the uncertainty on $f_{\text{DM}}$ is dominated by the uncertainty on the average mass of the PBH mass distribution.
The line of $f_{\text{DM}} = \langle M \rangle$ defines the direction in which the number density of PBHs is constant.
While the variance about the line is small (which is uncertainty on the number density of PBHs, $\psi_{\text{PBH}} N$), the length of that line is the uncertainty on the average PBH mass.
This uncertainty on the average mass directly reflects the resolution of our model.
The second and third mass bins, which dominate the PBH contribution to the overall population model, lie in this region of mass, and we are resolution limited smaller than that range of mass.

The far more intriguing result is the seeming inconsistency of $f_{\text{DM}}$ being zero, with a 5\% CI at 0.11.
This result is directly consistent with the Bayes factor calculation, as the strong evidence for the \model{NS Kick 150 + PBH} model indicates the extra complexity introduced by the PBH subpopulation is necessary to provide a better description of the data.
However, this result should be understood in its full context: this work only considered a finite number of models and only considered additional modeling in terms of PBH populations and adjustments to the astrophysical mass distribution.
Ultimately, our exploration of the model's performance in this section makes one thing clear: the PBH model is preferred primarily \emph{because of the modified velocity distribution}, giving it a better ability to describe the $t_E$-$\pi_E$ covariant space of observable microlensing parameters.
This feature is not unique to PBHs, although it is a generic component of any PBH theory.

\subsection{Discussion on Possible Astrophysical Systematics}

Within the realm of astrophysics, there are many sources of modeling systematics that could contribute to a total or partial explanation of these results.
Several of these are fully expected to exist in reality and were not included in our simulations, such as FFPs, BDs and binary systems (source and lens).
Our inference is expected to be reasonably robust to systematics at the low mass end (FFPs and BDs) due to the reasons outlined in Sec.~\ref{sec:forwardModeling}. 
That assessment can possibly be strengthened even more in light of the results in this section: the PBH distribution at lowest masses is inferred to be very low.
With the OGLE detection efficiency and event count so low at those timescales, the impact to the analysis was minimal.
However, it should be noted that possible FFP formation mechanisms might cause these planets to be ejected at high velocity from their host systems, thereby creating a population of fast moving, short timescale events, possibly complicating this (somewhat simplistic) extrapolation. 
This hypothesis would require dedicated modeling, as the exact mechanism from which a population of FFP is formed will drastically alter its distribution in $t_E$-$\pi_E$. 
Because of this, it is ill-advised to extrapolate the results of this work (based on restricted astrophysical and PBH modeling) to the case of FFPs with too much confidence, and future work is necessary to definitively assess this possibility.

Meanwhile, the question of binaries is a more open issue. The addition of binaries into a microlensing population tends to broaden the $t_E$ distribution and extend it at the long-$t_E$ end \citep{Abrams:2025}. This could possibly explain some of the effects missing in the preferred model; however, binaries would be expected to follow the same velocity distribution as the rest of the astrophysical population. In general, though, in OGLE and other microlensing surveys, only events with a strong binary signal (i.e. multiple peaks or significant asymmetry) have been fit with binaries models. From population synthesis modeling, there is likely an underlying population of binaries which are masquerading as PSPL events \citep{Abrams:2025}, and it is unclear exactly how these binary events being modeled as PSPL events affect parameter distributions and fit posteriors.

In this analysis we have begun to explore the inclusion of binaries and see improvement over the standard \textit{SukhboldN20} model, but there is substantial future work to be done to include them robustly. Beyond incorporating them into the better performing galactic models, there is also significant uncertainty in the exact binary model itself. We only test one configuration of binaries, the multiplicity fraction, number of companions per system, semi-major axis distribution, mass-ratio distribution, and eccentricity distribution all also have uncertainties. Binaries are an important piece of the puzzle that can often cause complications and reveal degenerate effects in modeling.

Furthermore, stellar streams are another possible source of confusion with some halo-distributed population, purely from the standpoint that a population of stars from a stellar stream would have a very different velocity distribution to the source distribution of velocities~\cite{}.
While this mechanism could possibly explain the excess of high velocity objects inferred to be in the catalog, a more careful treatment beyond the scope of this work is required. 
The number density and exact distribution in both velocity and space would be needed to accurately assess this population's ability to explain the discrepancy between the astrophysical models used in this work and OGLE data.

Beyond novel physics modeling and simulations, biases in the inferred distribution of $\pi_E$'s could be a symptom of non-astrophysical effects, such as systematics in the detection efficiency curves or unknown systematics in the modeling of the individual lightcurves.
Without full access to the OGLE image/lightcurve data, we cannot assess the robustness of their assessments of detection efficiency to things like parallax, binaries or finite-source effects, just to list a few possibilities which could cause a bias, already mentioned by \citet{Mroz2019}.
The accurate calculation of the detection efficiency is critical to studies such as this, and particularly in the regimes of low event counts where the inference on the \emph{underlying} population is tightly conditioned on understanding the selection bias of the detector or survey, as errors are drastically enhanced with low event counts.

In particular, the neglect of $\pi_E$ in the detection efficiency could impact our results.
We see from Fig.~\ref{fig:pbh_astro_piE_comp} that the \model{NS Kicks 150} and \model{NS Kicks 150 + Astro. Residual} models have more probability weight at higher parallax.
Compared to the parametric hierarchical inference (assumed to be more representative of the data itself, as it has no physical assumptions), this higher density at high parallax is disfavored, partially driving the evidence for the \model{NS Kicks 150 + PBH} model.
However, this is a comparison of the \emph{observed} distribution, and because we do not have detection efficiencies as functions of $\pi_E$, our forward models only consider selection effects via $t_E$ (only percolating to the $\pi_E$ distribution through correlations between the two parameters).
If the impact of $\pi_E$ on the detection efficiency is large enough, the astrophysical models might not be so disfavored, and the evidence for a fast moving, halo distributed population might be weakened.
However, we do believe our results are fairly robust to this particular systematic as the impact of parallax is minimized at low $t_E$ and maximized at high $t_E$.
Given the driver for the high parallax probability weight are the low mass, low timescale events, both in the astrophysical model and the two residual models, the impact of this systematic to these results should be minimal.

This is a crucial point not just for our results with regards to PBHs: the SOBH population could be much more susceptible to these systematics as they occupy the high $t_E$ regime. 
Our inference on the IFMR and SOBH abundance are tightly conditioned on the assessment of these detection efficiencies, as mentioned above.
To first order, the inferred abundance of SOBHs will scale linearly with any fractional change to the detection efficiency (i.e., dropping the SOBH abundance by 10\% with a 10\% shift in the average SOBH detection efficiency).

To resolve these issues concerning detection efficiency definitively, the detection efficiency as a function of $t_E$ and $\pi_E$ jointly would need to be undertaken using the same methods as ~\citet{Mroz2019}, but with a two dimensional grid instead of one.
Doing so in the full covariant space is critical, as selection effects as a function of $t_E$ and $\pi_E$ are certainly not statistically independent, and neglecting that covariance will lead to (possibly strong) bias.
As the data is not public, there is no recourse for this systematic with the OGLE survey data unless all the image data is made public by OGLE directly or the OGLE collaboration decides to undertake this work of their own accord. 

\section{Conclusions and Future Work}\label{sec:conclusions}

In this work, we have analyzed the OGLE-IV survey data using hierarchical Bayesian inference to better understand the population of lenses towards the Galactic Bulge.
We begin by assessing the performance of a targeted (but not exhaustive) suite of astrophysical models meant to probe specific and largely uncertain modeling considerations, including the Galactic Bulge orientation and dynamics, binary system formation and the physics of NS and SOBH formation (both in their mass distribution via the choice of IFMR and kick velocities).

Through this initial model exploration, we find that microlensing surveys are less sensitive to Galactic Bulge orientation\footnote{This is with the caveat that we do not utilize physical modeling for the rate of microlensing events, only relying on simulation to produce the distributions for the observable parameters $t_E$ and $\pi_E$.}.
However, microlensing does seem to be an effective probe of Galactic Bulge dynamics, favoring models with slower bar patternspeed.
Both the Galactic Bulge models with lower bar patternspeeds were heavily favored over the Galactic Bulge model with a higher bar patternspeed, despite having different spatial configurations for the bar.

We also consider quantitatively for the first time the presence of binary systems in the hierarchical inference of microlensing surveys. Although preliminary findings show the model is not the best performing, including binaries does improve the model evidence over the standard \model{SukhboldN20}. Further, more extensive investigations could exhaustively search the model space for the best combination of population modeling effects.
We note that the fact that binaries do not perform the best does not necessarily lead to the conclusion that binaries are not needed generally, as we only considered binaries on top of the \model{SukhboldN20} IFMR, the \model{Galaxia v${}_3$} galactic model and NS natal kicks of $350$ km/s, none of which were favored by the data. 
Future work could study this effect more in depth, addressing some of the nuances of mapping binary lensing parameters to the PSPL-focused observable parameter space.
Our initial effort used a simplistic processing of the events, assuming rough cuts to remove things that are not sufficiently PSPL-like to match OGLE-style selection criteria.
Beyond this filtering, the exact meaning of $t_E$ becomes ill-defined for binary events, and future work is needed to better understand the mapping of PSPL-specific event parameters (with and without parallax) to the physical parameters of binary systems. 
This definition is critical as our framework assumes an unbiased, accurate description of the distributions of event modeling parameters, but it is poorly understood, as the mapping from PSPL to binary lensing parameters and its implication on this analysis has only been studied preliminarily~\citep{Abrams:2025}.

With regards to NS and SOBH formation, we find that the \citet{Spera2015} IFMR is the most favored model of the three theories considered in this work, followed by \citet{Raithel2018} and finally \citet{Sukhbold2014}.
Given the preceding work of \citet{Rose2022}, this is believed to be the result of the wider mass distribution of \citet{Spera2015} in both the NS and SOBH, which seems to be favored by the data.
For the NS kick velocities, microlensing again seems to be an effective probe.
The most favored model in the entire suite of astrophysical models assumed an average NS natal kick velocity of 150 km/s, which is 200 km/s slower than what has been assumed in the past for modeling the Galactic Bulge.

With a best performing astrophysical model, we considered two additional models meant to assess systematic bias and hint at additional modeling that might be needed to explain the data.
Both models implement a flexible mass distribution into the modeling/inference pipeline, but one does so by introducing a subpopulation of PBH lenses while the other adds a subpopulation of astrophysical lenses.
The model incorporating PBHs into the population drastically outperforms both of the other contending models, with a Bayes factor of $\ln\mathcal{B}^{\text{\model{NS Kick 150 + PBH}}}_{\text{\model{NS Kick 150}}} = 20.23$.
We then present evidence, based on the one dimensional marginal $t_E$ and $\pi_E$ posterior predictive distributions as well as the fully covariant $t_E$-$\pi_E$ model predictions and inferred parameters from \citet{Golovich2022}, that this increase in model performance is primarily due to the unique velocity distribution. 
The PBH velocity distribution is starkly different from the average velocity of lenses and sources in the astrophysical-only galactic model.
This difference allows for a better description of a large number of events at low $t_E$ and low $\pi_E$, which are weakly explained by the contending astrophysical only and astrophysical residual models.

Acknowledging that PBHs are not the only explanation for this effect on the observable parameter space, we move on to consider the PBH more seriously as one possible contender.
By exploring the final posterior inference on the hyperparameters of the PBH model, we infer that this population does not favor a low mass ($\lesssim 0.08 M_{\odot}$) component, instead being heavily skewed towards PBH lenses in the $0.08$ - $2.9 M_{\odot}$ range.
The inferred mass distribution largely overlaps with the Stellar population (noting that an overlapping mass distribution does not necessarily lead to fully degenerate predictions for $t_E$ and $\pi_E$).

We then go on to assess what fractions of DM would be compatible with this inferred population of PBH lenses, arriving at a $95\%$ upper limit on the DM fraction of $0.30$. 
Consistent with the Bayes factor result for this model, we also find a $5\%$ lower bound on $f_{\text{DM}}$ at $0.11$, suggesting a strong preference for this additional modeling component with a disparate velocity distribution compared to the general galactic bulge velocity distribution.
As discussed in the results above, we go on to briefly discuss other physical processes, populations and survey characterization which may also explain these features without resorting to a hypothetical and exotic subpopulation.
In the modeling realm, these possibilities include FFPs, BDs, binaries and stellar streams as possible systematics.
With regards to uncertainty quantification for the survey characterization, we note that more detailed calculation of the detection efficiency of OGLE for both $t_E$ and as a function of $\pi_E$ incorporating binaries and parallax would help to confirm or challenge this work. 

PBHs are also not the only DM candidates which can generate a microlensing signature. In particular, in dissipative DM models, DM can emit dark radiation and cool efficiently. Dissipative DM models are generally constrained to be ${< 10}$\% of the total DM density, lest they overly disrupt galaxy formation~\citep{FAN2013139, PhysRevD.89.063517, chacko2018cosmologicalsignaturesmirrortwin}. However, a small component of dissipative DM can form a dark disc~\citep{FAN2013139, PhysRevLett.110.211302}, within which further collapse produces microlensing dark compact objects~\citep{Winch_2022}. Objects in the dark disc would trace the stellar disc and thus would have a low density towards the LMC, evading OGLE microlensing constraints. However, their velocity dispersion could be similar to that of the halo, depending on the amount of dissipation. A population of dark compact objects in a dark disc thus presents an explanation for the population of anomalous lenses we observe towards the galactic bulge, without being in conflict with the OGLE constraints.

In future work, we expect these results to be improved and cross-checked with the up-coming Roman Space Telescope \citep[e.g.,][]{DeRocco2024, Fardeen2023}. 
In particular, precision parallax measurements at a population-level would help to test some of these conclusions in particular, as we are largely being driven by the upper limit of the parallax distribution. 
This is in addition to all the other benefits Roman Space Telescope will bring, including higher detection rates and precision photometry, as well the possibility for large-scale astrometry efforts.
In the meantime, applying this analysis with the more complicated models to the LMC/SMC data from OGLE might provide additional insights into the allowable parameter space, as past works only considered point estimates in the galactic model space and monochromatic mass functions.

Of course, this work only began the process of confronting current galactic and lens population modeling efforts with data.
Future work can continue to hone in on the astrophysical modeling space, refining the inference performed here to tailor it to astrophysical model inference with more targeted goals.
Furthermore, additional modeling software and frameworks could be added to future work, expanding our inference to some of these populations not currently included by \texttt{PopSyCLE}, such as FFPs and BDs.

\section*{Acknowledgments}
This work was performed under the auspices of the U.S. Department of Energy by Lawrence Livermore National Laboratory under Contract DE-AC52-07NA27344. The document number is \IMRELEASENO{}. This work was supported by the LLNL-LDRD Program under Project 22-ERD-037.  N.S.A. and J.R.L. acknowledge support from the National Science Foundation under grant No.~1909641 and the Heising-Simons Foundation under grant No.~2022-3542. SB acknowledges funding from NASA ATP 80NSSC22K1897. MFH is supported by a NASA FINESST grant No. ASTRO20-0022, the Leinweber Foundation and DOE grant DE-SC0019193. \\

\vspace{5mm}

\software{This research has made use of NASA's Astrophysics Data System Bibliographic Services. NumPy \citep{Harris2020}, SciPy \citep{Virtanen2020}, Matplotlib \citep{Hunter2007}, corner \citep{corner}, emcee \citep{Foreman-Mackey2013}, scikit-learn \citep{sklearn_api}, Singularity \citep{Kurtzer2017, kurtzer2021}, Docker \citep{merkel2014}, Numpro \citep{Phan2019}}, PopSyCLE \citep{Lam2020}, SPISEA \citep{Hosek2020}

\bibliography{refs}{}
\bibliographystyle{aasjournal}

\end{document}